\documentclass[reprint,aps,prb,twocolumn,nobibnotes]{revtex4-1}
\usepackage[utf8]{inputenc}
\usepackage{amsmath}
\usepackage{physics}
\usepackage{braket}
\usepackage{amssymb}
\usepackage{graphicx}
\usepackage{xcolor}
\usepackage[most]{tcolorbox}
\usepackage[unicode=true, breaklinks=false, pdfborder={0 0 1}, backref=false, colorlinks=true, linkcolor=blue, urlcolor=blue, citecolor=blue]{hyperref}
\setcitestyle{numbers,square}

\begin{document}   

\title{Nonreciprocal surface magnetoelastic dynamics} 
\author{Tao Yu} 
\email{tao.yu@mpsd.mpg.de}
\affiliation{Max Planck Institute for the Structure and Dynamics of Matter, Luruper Chaussee 149, 22761 Hamburg, Germany} 

 \begin{abstract}
 Motivated by recent experiments, we investigate the nonreciprocal magnetoelastic interaction between the surface acoustic phonons of dielectric non-magnetic substrates and magnons of proximity nanomagnets. The magnetization dynamics exerts rotating forces at the edges of the nanomagnet that causes the nonreciprocal interaction with surface phonons due to its rotation-momentum locking. This coupling induces the nonreciprocity of the surface phonon transmission and a nearly complete phonon diode effect by several (tens of) magnetic nanowires of high (ordinary) magnetic quality. Phase-sensitive microwave transmission is also nonreciprocal that can pick up clear signals of the coherent phonons excited by magnetization dynamics. Nonreciprocal pumping of phonons by precessing magnetization is predicted using Landauer-B\"uttiker formalism.

 \end{abstract}
\date{\today}

\maketitle

\section{Introduction}
Efficient transfer of spin information
among different entities is the prerequisite to achieve long-range spin transport in
spintronics \cite{spintronics,spintronics_RMP}. The spin diffusion
length can be of micrometers in two-dimensional electron gas \cite{Marie} and even longer in graphene \cite{graphene_diffusion,graphene_RMP}. Long-wavelength dipolar spin waves in the magnetic insulator---yttrium iron
garnet (YIG)---can even travel over
centimeters \cite{YIG_centimeter}, but they suffer from a low group velocity; exchange spin waves has a large group velocity but their lifetime is shorter \cite{magnonics1,magnonics2,magnonics3,magnonics4}. Recent studies showed that bulk phonons in the insulator gadolinium gallium garnet (GGG) can couple two YIG
magnetic layers over millimeters \cite{phonon_long1,phonon_long2,Simon,Kruglyak}, raising the possibility of using phonon currents to
transfer spin information in non-magnetic insulators. 
The surface (Rayleigh) acoustic waves (SAWs), known as excellent sources to pump spin waves via acoustic spin pumping \cite{acoustic_pumping1,acoustic_pumping2,acoustic_pumping3,acoustic_pumping4,acoustic_pumping5,acoustic_pumping6,acoustic_pumping7}, can propagate a longer distance with a larger group velocity \cite{acoustic_millimeter,SAW_book1} and thus is promising to transport spin information. 

Very recently, the nonreciprocal surface phonon transmissions were observed when the phonons pass through the ultrathin extended ferromagnetic films in proximity to the piezoelectric substrate \cite{Xu,DMI} and are explained by the magnetorotation or/and magnetoelastic couplings \cite{paper1976}. These indicate that the interaction between magnons and surface phonons is nonreciprocal (or chiral when emphasizing the symmetry), i.e., the magnons in the magnets can dominantly couple the traveling surface phonon propagating in one direction. The inverse process of acoustic pumping---the nonreciprocal pumping of phonon by magnetization dynamics---has not yet been experimentally reported and was theoretically considered by us \cite{Xiang}. There, interference with dynamical phase shift $\pi$ between two remote magnetic nanowires that couples with the phonon reciprocally is responsible rather than a direct nonreciprocal magnon-phonon coupling in the presence of one magnet \cite{Xiang}.

Magnons hold chirality by their anticlockwise rotation and are revealed recently to be able to nonreciprocally couple with various quasiparticles or devices. The long-range dipolar interaction emitted from the excited nanomagnet can chirally couple with the traveling magnons of extended films as its rotation direction is locked to its momentum \cite{magnon_magnon1,magnon_magnon2,magnon_magnon4,magnon_magnon3,magnon_trap_exp}.  The microwaves show polarization-momentum locking when confined by the waveguide \cite{waveguide1,waveguide2,waveguide3,waveguide4}, cavity \cite{Tang,cavity} or antenna \cite{antenna1,antenna2,antenna3}, which were employed to realize the nonreciprocal magnon-photon coupling. With chiral coupling, the unidirectional traveling waves are excited by the nearby magnet in half space \cite{magnon_magnon4,magnon_magnon3,magnon_trap_exp}. The evanescent dipolar field or microwaves can realize non-contact (and chiral) spin pumping to the nearby conductors \cite{magnon_electron}. The traveling waves mediate a long-range nonreciprocal interaction between remote magnets and the spin accumulates at the edge of magnets by the non-Hermitian skin effect \cite{skin0,skin1,skin2}.  Interference effect in nonreciprocal systems can directionally amplify or trap the traveling waves \cite{magnon_trap,Canming}.

The surface acoustic waves exhibit rotation-momentum locking as well \cite{SAW_book1,SAW_book2}, from which their nonreciprocal coupling with magnons may be understood universally. In this work, we study the surface magnetoelastic coupling in the spin mechanical system and formulate the nonreciprocal dynamics \cite{magnon_magnon4,waveguide4,magnon_electron} via the Green function method \cite{Haug,Mahan,Fetter,Abrikosov}. Rather than considering the extended magnetic film in which the edge effect is marginal \cite{Xu,DMI,paper1976}, we focus on the thin nanomagnets with dominant edge effect. We show the uniform magnetization dynamics exerts rotating forces at the edges of the nanomagnet that causes the nonreciprocal interaction by the rotation-momentum locking of surface phonons. As the magnons dominantly couple with surface phonon propagating in one direction, the surface phonon transmission is nonreciprocal. We also propose to detect the nonreciprocal coupling by the phase-sensitive microwave transmission \cite{magnon_magnon3,magnon_trap_exp}: the microwaves can excite the magnetization of one magnet that pumps the unidirectional phonon propagation, which can in turn excite another magnet, above which the signal is picked up by the radiated microwaves. This method can detect clear signals of the excited coherent phonons and their group velocity. With the nonreciprocal magnetoelastic coupling, we predict the unidirectional pumping of phonons by ferromagnetic resonance (FMR).

This paper is organized as follows. We model and calculate the coupling between magnon and surface phonon
in Sec.~\ref{sec:coupling}. In Secs.~\ref{sec:scattering_general} and \ref{sec:transmission}, the phonon and microwave transmissions are addressed. The directional pumping effect is discussed by the Landauer-B\"uttiker formalism in Sec.~\ref{sec:Landau_Buttiker}. We summarize and discuss the results in Sec.~\ref{sec:summary}.

\section{Nonreciprocal magnon-phonon interaction}
\label{sec:coupling}
\begin{figure}[ht]
	\centering
	\includegraphics[width=8.2cm]{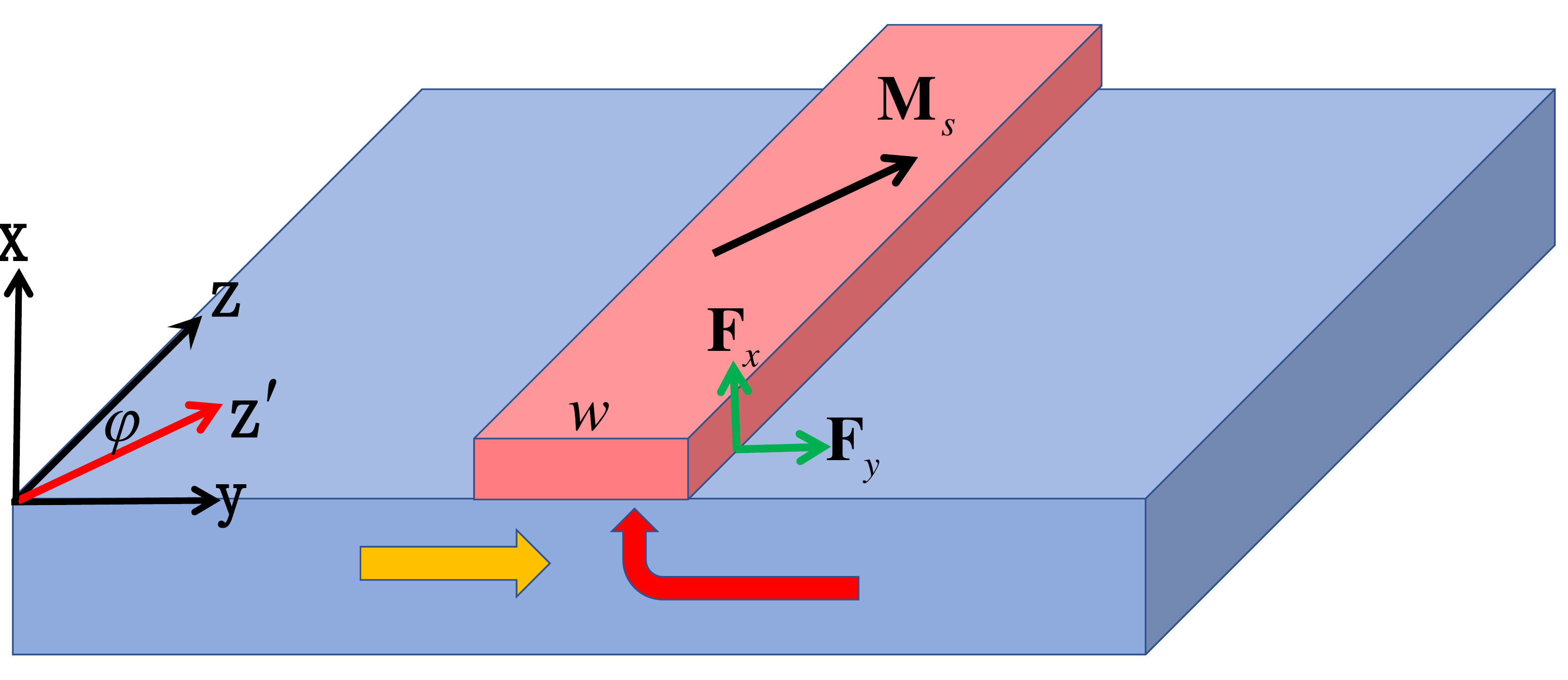}
	\caption{Magnetic nanowires on top of a dielectric substrate. A strong magnetic field is applied to change the magnetization direction, parameterized by the angle $\varphi$ relative to the wire direction. With a precessing magnetization, rotating forces emerge at the left and right edges of the nanowire (illustrated by $F_{x,y}$ at the right edge). The orange and red thick arrows indicate the nonreciprocal propagation of the SAWs.}
	\label{fig:phonon}
\end{figure}

We consider magnetic nanowires of width $w$ and thickness $d$ ($\ll w$) on top of dielectric substrates as illustrated in Fig.~\ref{fig:phonon}, focusing on \textit{both} the magnetoelastic and magnetorotation couplings between them. The dielectric substrate is assumed to be semi-infinite; it is usually not magnetic, since non-magnetic substrates are used in the recent experiments, which can be GGG \cite{phonon_long1}, MgO \cite{Xu} or Pt \cite{DMI}. We assume the thickness $d$ [$O(10~{\rm nm})$] is much smaller than the decay length or wavelength $\lambda$ of the SAWs ($>100~{\rm nm}$), while the width $w$ is comparable to $\lambda$.  Experimentally, such a geometry with thin Cobalt or Nickel nanowires on top of a thin YIG film was used to  realize the pumping of short-wavelength spin waves \cite{magnon_magnon3,magnon_trap_exp,Jilei_Ni}. We restrict the magnetization to be parallel to the substrate surface but allow an angle between it and the nanowire $\hat{\bf z}$-direction \cite{Xu}. To this end, we assume a sufficiently strong magnetic field $-H_0\hat{\bf z}'$ is applied, with an angle $\varphi$ between ${\bf z}'$ and the wire ${\bf z}$-direction, to saturate and control the direction of the wire magnetization with an equilibrium component $\sim -M_s\hat{\bf z}'$ and transverse components $m_{x'}{\bf x}'+m_{y'}{\bf y}'$ (see Appendix). Since the spins of electron are opposite to the magnetization, the spins are (nearly) parallel to $\hat{\bf z}'$. 

In our configuration, only the SAWs couple efficiently with the nanomagnets by their surface nature in that sufficiently thin nanowire does not affect the substrate strongly and their interaction can be treated perturbatively. We employ the quantum description that allows to study the dynamics by Green function technique but also explain our findings by classical picture.
The Hamiltonian consists of the elastic $\hat{H}_{\mathrm{e}}$, magnetic $\hat{H}_{\mathrm{m}}$, and their coupling $\hat{H}_{\mathrm{c}}$. In Appendix~\ref{appendix}, we quantize $\hat{H}_{\mathrm{e}}=\sum_k\hbar \omega_k\hat{b}_k^{\dagger}\hat{b}_k$ in terms of phonon operator $\hat{b}_k$ with $k=k_y$ and $\hat{H}_{\mathrm{m}}=\hbar \omega_{\rm K}\beta_{l}^{\dagger}\hat{\beta}_l$ with $\hat{\beta}_l$ being the magnon operator in the $l$-th nanomagnet. Here, $\omega_k$ and $\omega_{\rm K}$ are the frequencies of the phonon of momentum $k$ and Kittel magnon, respectively. $\hat{H}_{\mathrm{c}}=\hat{H}^{(a)}_{\mathrm{c}}+\hat{H}^{(b)}_{\mathrm{c}}$ is contributed by the magnetoelastic coupling ($a$) and magnetorotation coupling ($b$), as addressed below.

\subsection{Magnetoelastic coupling}
The magnetoelastic coupling Hamiltonian generally depends on the crystal symmetry of the material \cite{Landau}. Here we adopt the simplest form that describes a wide class of material \cite{Kittel_old,parameters,Simon,Xu,DMI}, which may be written as \cite{Landau,Simon,Kittel_old}
\begin{equation}
\hat{H}^{(a)}_c=\frac{1}{M_s^2}\int d{\bf r}\left(B_{\parallel}\sum_iM_i^2\varepsilon_{ii}+B_{\perp}\sum_{i\neq j}M_iM_j\varepsilon_{ij}\right),
\label{magnetoelastic}
\end{equation}
where $B_{\parallel}$ and $B_{\perp}$ are the magnetoelastic constants, and $\varepsilon_{ij}=(\partial_j u_i+\partial_i u_j)/2$ denotes the strain tensor in terms of the displacement field $u_i({\bf r})$.
For the Rayleigh SAWs propagating perpendicular to the wire with momentum ${\bf k}\parallel\hat{\bf y}$, there only exists $(u_x(x,y),u_y(x,y))$ and only $\varepsilon_{xx}$, $\varepsilon_{yy}$ and $\varepsilon_{xy}$ are non-vanishing. The Hamiltonian can be linearized when the temperature is far below the Curie temperature. Considering the coupling to the uniform Kittel mode in the nanowire, Eq.~(\ref{magnetoelastic}) is linearized to be 
\begin{align}
\nonumber
\hat{H}^{(a)}_c&=\frac{2\sin\varphi}{M_s}\int d{\bf r}\left(B_{\parallel}\cos\varphi m_{y'}\varepsilon_{yy}+B_{\perp} m_{x'}\varepsilon_{xy}\right)\\
\nonumber
&\simeq \frac{2\sin\varphi\cos\varphi}{M_s} {B_{\parallel}Ld}m_{y'}\left(u_y|_{\frac{w}{2}+y_l}-u_y|_{-\frac{w}{2}+y_l}\right)\\
&+\frac{\sin\varphi}{M_s}B_{\perp}Ldm_{x'}\left(u_x|_{\frac{w}{2}+y_l}-u_x|_{-\frac{w}{2}+y_l}\right),
\label{eqn:coupling}
\end{align}
where $y_l$ is the center coordinate of the $l$-th nanomagnet and $L$ is the length of the nanowire.
Here we have assumed that the nanowire is sufficiently thin such that the displacements at its top and bottom surface are identical and hence have no net contribution to the magnetoelastic coupling.

Classically, we obtain the forces, by $F_{x,y}({\bf r})=\delta H_c/\delta u_{x,y}({\bf r})$ \cite{Simon,Xiang}, of the $l$-th nanowire at the right edge ($y=w/2+y_l$)  
\begin{align}
\nonumber
&F_y|_{y=\frac{w}{2}+y_l}=2\sin\varphi\cos\varphi B_\parallel Ldm_{y'}/M_s,\\ 
&F_x|_{y=\frac{w}{2}+y_l}=\sin\varphi B_\perp Ld m_{x'}/M_s,
\end{align}
and at the left edge ($y=-w/2+y_l$) 
\begin{align}
\nonumber
&F_y|_{y=-\frac{w}{2}+y_l}=-2\sin\varphi\cos\varphi B_\parallel Ldm_{y'}/M_s,\\ 
&F_x|_{y=-\frac{w}{2}+y_l}=-\sin\varphi B_\perp Ld m_{x'}/M_s.
\end{align}
The generated forces are opposite at the two edges of the wires.
There are generally both $x$ and $y$ components in the forces that are rotating when the magnetization $m_{x'}$ and $m_{y'}$ rotate (Fig.~\ref{fig:phonon}). Although they are not circularly polarized even when the magnetization are, the elliptically polarized forces bring chirality in the mechanics. When $\varphi=\pi/2$ ($\varphi=0$) with the magnetization perpendicular (parallel) to the wire, the force becomes linearly polarized (vanish), recovering our previous results \cite{Xiang}. Since the rotation direction of the SAWs is locked to their momenta [see Eq.~(\ref{eqn:SAW_profile})], we expect the coupling between magnon and phonon is nonreciprocal. 

We note that although in the classical description, the total free energy Eq.~(\ref{eqn:coupling}) and the total forces depend on the length of the wire $L$, the excited SAW amplitude is independent of $L$ since only the force density, i.e. the stress, plays a role \cite{Simon}. A detailed description of the forces and how they excite the SAWs, for the perpendicular configuration, are given in our previous work \cite{Xiang}. We showed there that the forces arise at the left and right boundaries of a wire and exert the stress on the dielectric substrate that excite the SAWs. The excited SAWs then propagate away from the regions of the wires. Here we focus on the quantum description that recovers and extends the results from classical treatment.

 \subsection{Magnetorotation coupling}
 Magnetocrystalline and shape anisotropies can contribute to the magnetorotation coupling in terms of antisymmetric tensor $\omega_{ij}=(\partial_ju_i-\partial_iu_j)/2$ \cite{Maekawa,magnetorotation_1997}. Here we consider the uniaxial anisotropy, for simplicity. The coupling between magnon and surface phonon generally depends on the uniaxial direction relative to the wire $\hat{\bf z}$-direction. It vanishes when the easy axis is along the wire since $\omega_{xz}$ and $\omega_{yz}$ vanish for SAWs propagating along the $\hat{\bf y}$-direction. 
 
 We first consider the perpendicular anisotropy with the easy axis along the $\hat{\bf x}$-direction \cite{Xu}, yielding the Hamiltonian
  \[
  \hat{H}_c^{(b)}=-\frac{2K_1}{M_s^2}\int d{\bf r}M_x({\bf r})\big[M_y\omega_{yx}({\bf r})+M_z\omega_{zx}({\bf r})\big],
  \]
  where $K_1$ is contributed by the uniaxial anisotropy field. This Hamiltonian is linearied to be
  \begin{align}
  \nonumber
  \hat{H}_c^{(b)}&=\frac{K_1}{M_s}\sin\varphi m_{x'}\left(\frac{\partial u_x}{\partial y}-\frac{\partial u_y}{\partial x}\right)\\
  &=\frac{\sin\varphi}{M_s}K_1L dm_{x'}\left(u_x|_{\frac{w}{2}+y_l}-u_x|_{-\frac{w}{2}+y_l}\right),
  \label{magnetorotation_coupling}
  \end{align}
  contributing a force perpendicular to the substrate surface at the edge of the nanowire. Comparing with Eq.~(\ref{eqn:coupling}), we conclude that including the perpendicular anisotropy here shifts $B_{\perp}$ to $\tilde{B}_{\perp}=B_{\perp}+K_1$. We then similarly address the case with the easy axis along the $\hat{\bf y}$-direction. We find $\tilde{B}_{\perp}=B_{\perp}-K_1$.
 
 \subsection{Coupling Hamiltonian}
 The polarization of the rotating forces follows that of the Kittel magnon. The Kittel mode is linearly polarized under the weak applied magnetic field and the induced force is not rotating. As only a circularly polarized magnon favors the nonreciprocity, we assume that a large magnetic field $H_0$ is applied such that the magnon is circularly polarized in the thin wire (see  Appendix~\ref{appendix}).
 By substituting the magnetization operator Eq.~(\ref{eqn:magnon_operator}) and displacement-field operator Eq.~(\ref{eqn:phonon_operator}) into Eqs.~(\ref{eqn:coupling}) and (\ref{magnetorotation_coupling}), the coupling Hamiltonian becomes
 \begin{equation}
 \hat{H}_c=\hbar\sum_l\sum_{k}g_{l}(k)\hat{\beta}_l^{\dagger}\hat{b}_{k}+\mathrm{H.c.},
 \end{equation}
 with the coupling constant
 \begin{align}
 \nonumber
 g_{k,l}&=i\sin\varphi\sqrt{\frac{\gamma}{M_s\rho c_r}}\sqrt{\frac{d}{w}}\sin\left(\frac{kw}{2}\right)e^{iky_l}\xi_P\\
 &\times\left(\tilde{B}_\perp-\cos\varphi B_\parallel\mathrm{sgn}(k)\frac{1+b^2}{a}\right).
 \label{eqn:coupling_strength}
 \end{align}
 Here, $-\gamma$ is the gyromagnetic ratio of electron; $\rho$ and $c_r$ are the density of the dielectric substrate and the group velocity of the surface phonon, respectively; $\xi_P$, $a$ and $b$, given by Eqs.~(\ref{eqn:ksi_P}) and (\ref{eqn:alpha_beta}), are determined by elastic properties. The coupling constant depends on the sign of momentum and generally show the nonreciprocity with $|g_{|k|}|\ne|g_{-|k|}|$.
 We see that the coupling tends to vanish when $\varphi=0$, while when $\varphi=\pi/2$, there is no chirality as $|g_{|k|}|=|g_{-|k|}|$ \cite{Xiang}. Considering $\varphi\in (0,\pi/2)$, the complete chirality arises when $\tilde{B}_{\perp}=\cos\varphi_c B_{\parallel}(1+b^2)/a$ such that $g_{|k|}=0$, implying the critical angles $\varphi_c$ satisfy
 \begin{align}
 \cos\varphi_c=\frac{\tilde{B}_{\perp}}{B_{\parallel}}\frac{a}{1+b^2}.
 \end{align}
 As the critical angle is only determined by the basic material parameters and is not related to geometry parameters and the wave number of the phonons, it is fixed with the chosen material. This allows to choose optimal material for applications.
 Such conclusion agrees with the classical description in which the rotating force $|F_x|=|F_y|(1+b^2)/(2a)$ matches the phonon chirality. Without the magnetoelastic coupling, the magnetorotation coupling itself cannot cause nonreciprocity in the thin wire configuration.
 
The nonreciprocity is sensitive to the relative magnitude of $\tilde{B}_{\perp}$ and $B_{\parallel}$, which are usually  extracted experimentally. 
When the elastic substrate is GGG, ${a}/({1+b^2})=0.76$ \cite{Xiang}. For the Cobalt nanowire, $\tilde{B}_\perp\approx -9.2\times 10^6$~${\rm J/m^3}$ and $B_\parallel\approx 7.7\times 10^6$~${\rm J/m^3}$; for the Nickel nanowire, $\tilde{B}_\perp\approx B_\parallel=1.3\times10^7$~${\rm J/m^3}$ \cite{parameters}. All these two materials can achieve a complete nonreciprocity with $\cos\varphi_c \sim \pm{a}/({1+b^2})$. Nevertheless, complete nonreciprocity cannot be achieved for YIG$|$GGG  with a small anisotropy since $\tilde{B}_{\perp}\approx 2B_{\parallel}=6.96 \times 10^{5}$~${\rm J/m^3}$ \cite{Simon} leads to $\cos\varphi_c>1$. Therefore, magnetoelastic coupling may not always promise a complete nonreciprocity, different from the couplings of magnon with other quasiparticles \cite{magnon_magnon4,waveguide4,magnon_electron}.

\begin{figure}[ht]
 	\centering
 	\includegraphics[width=6.8cm]{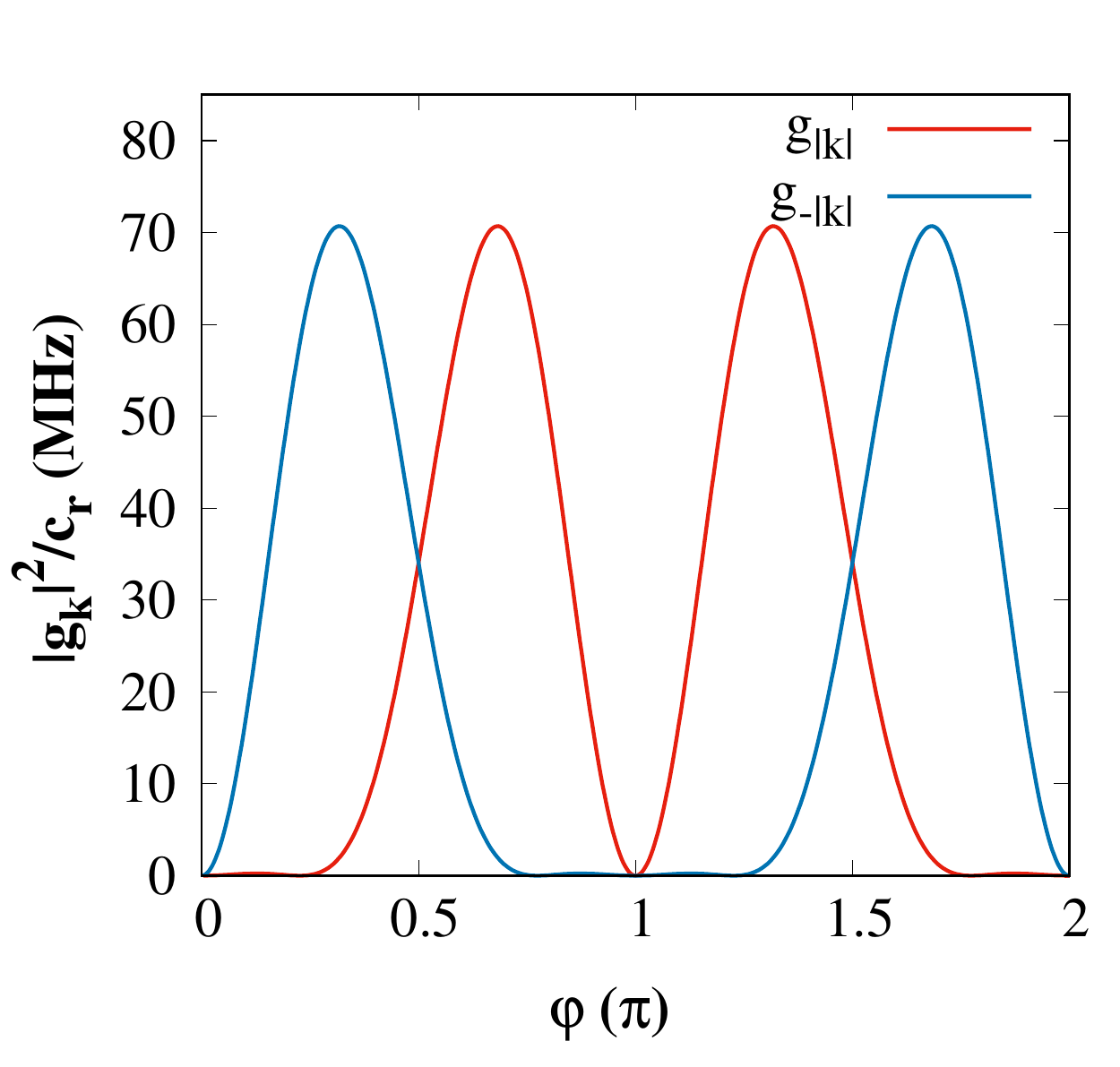}
 	\caption{Dependence on angle $\varphi$ of the  coupling strength $g_{\pm |k|}$ for Ni nanowire on GGG substrate.  Complete nonreciprocity arises at the critical angles $\varphi_c\approx 0.73 \pi$ and $1.27\pi$  with $g_{-|k|}=0$. The material parameters are given in the text. }
 	\label{fig:phonon_chirality}
 \end{figure}

In Fig.~\ref{fig:phonon_chirality}, we illustrate the dependence of the normalized coupling strength on the angle $\varphi$ for a Ni nanowire on the GGG substrate. We adopt Ni nanowire of magnetization $\mu_0M_s=0.5$~T, width $w=250$~nm and thickness $d=30$~nm \cite{magnon_magnon3,magnon_magnon1,Jilei_Ni}. The velocity of surface phonon $c_r=3271.78~{\rm m/s}$ in GGG. In the calculation, the FMR is fixed to be $2\pi\times$20~GHz by tuning the magnetic field around $\mu_0H_0=$1~T. Complete nonreciprocity arises at the critical angles $\varphi_c\approx 0.73 \pi$ and $1.27\pi$ with $g_{-|k|}=0$. 

 \section{Phonon diode effect}
 \label{sec:scattering_general}

 The nonreciprocal couplings are detectable by transmission spectra of phonon as the scattering cross section, for the phonon with opposite momenta, can be different, i.e., a diode effect for surface acoustic phonon \cite{Xu,DMI}. The phonon transmission should be tunable by the number of proximity nanomagnet. Therefore, not restricting to one nanomagnet, we generally calculate the phonon scattering matrix in the presence of $N$ parallel magnetic nanowires by using the scattering theory (Sec.~\ref{phonon_scattering}) \cite{scattering_PRB,scattering_PRE,Mahan}. The distance of the neighbouring wires is assumed to be much larger than the wire width such that the dipolar interaction between wires can be safely disregarded. In this case, every wire can be treated to be isolated. With many wires, the magnons in different wires interact with each other through virtual exchange of surface phonon; they form collective modes. We show that the phonon reflection are determined by these collective modes (Sec.~\ref{collective_modes}).  

\subsection{Phonon scattering matrix}
\label{phonon_scattering}

 The scattering amplitude between state $\ket{k}$ to state $\ket{k'}$ is calculated by the $T$-matrix \cite{scattering_PRB,scattering_PRE,Mahan}
 \begin{align}
 \nonumber
 T_{k'k}&=\langle k'|\hat{T}|k\rangle\\ &=\delta_{kk'}+\frac{1}{\omega_{k}-\omega_{k'}+i\eta_{k'}}\sum_{ll'}g^*_{k',l}
 G_{N,ll'}(\omega_{k})g_{k,l'},
 \label{eqn:T_matrix}
 \end{align}
 where $\eta_{k'}$ is the damping broadening of phonon.
 The summation on the magnet index $\{l=1,2,\cdots,N\}$ implies that the scattering from state $\ket{k}$ to state $\ket{k'}$ experiences all possible scattering paths such that the phonon transmission results from the net interference. Note that the magnets are not isolated as they interact with each other via the exchange of phonons. Also, different from the static scatters, the magnets are dynamical such that they absorb and emit the phonons. All these dynamics is encoded in the magnon Green function $G_{N}|_{ll'}$ that stands for the propagator of magnon from wire $l'$ to $l$.  It is calculated to be \cite{Mahan,Fetter,Haug}
 \begin{equation}
 \left(G^{-1}_{N}(\omega)\right)_{ll'}=(\omega-\tilde{\omega}_{\rm K})\delta_{ll'}-\Sigma_{ll'}(\omega),
 \label{eqn:green_function}
 \end{equation}
 where $\tilde{\omega}_{\rm K}=\omega_{{\rm K}}-i\alpha_G\omega_{{\rm K}}$ with the Gilbert damping parameterized by the Gilbert coefficient $\alpha_G$, and $\Sigma$ is the self-energy matrix of magnons due to their collective interactions with phonons with elements 
 \begin{align}
 \Sigma_{ll'}(\omega)=\sum_{k'} \frac{g_{k',l}g^*_{k',l'}}{\omega-\omega_{k'}+i\eta_{k'}}.
 \end{align}
 The Green function is the basis for calculating the coherent and incoherent dynamics below.

  When $\eta_{k}\rightarrow 0_+$ for the high-quality elastic substrate such as GGG, these elements are calculated to be
 \begin{align}
 \nonumber
 &\Sigma_{ll}(\omega)=-\frac{i}{2c_r}\left(|g_{k_*,l}|^2+|g_{-k_*,l}|^2\right)=-i\Gamma_l(\omega),\\
 \nonumber
 &\Sigma_{l<l'}(\omega)=-\frac{i}{c_r}\tilde{g}_{k_*,l}\tilde{g}^*_{k_*,l'}e^{ik_*|y_l-y_{l'}|}=-i\Gamma_{R,ll'}(\omega),\\
 &\Sigma_{l>l'}(\omega)=-\frac{i}{c_r}\tilde{g}_{-k_*,l}\tilde{g}^*_{-k_*,l'}e^{ik_*|y_l-y_{l'}|}=-i\Gamma_{L,ll'}(\omega),
 \end{align}
 where $k_*=\omega/c_r+i\eta_k/c_r\rightarrow \omega/c_r$ and $\tilde{g}_{k,l}=g_{k,l}e^{-iky_l}$. 
 The diagonal elements $\Sigma_{ll}(\omega_{k})$ represent the broadening of magnons modes by pumping phonons. While the off-diagonal self-energies $\Sigma_{l\ne l'}(\omega_{k})$ imply that the magnets interact with each other mediated by the surface phonons.
 Thus, the magnon Green function is represented by $G_N(\omega)=\big(\omega-{\cal H}_N(\omega)\big)^{-1}$, where the matrix
 \begin{align}
 \nonumber
 {\cal H}_N(\omega)&\equiv \tilde{\omega}_{\rm K}\\
 &- i\begin{pmatrix}
 \Gamma_{1}(\omega)&\Gamma_{L,21}(\omega)&\cdots&\Gamma_{L,N1}(\omega)\\
 \Gamma_{R,12}(\omega)&\Gamma_{2}(\omega)&\cdots&\Gamma_{L,N2}(\omega)\\
 \cdots&\cdots&\cdots&\cdots\\
 \Gamma_{R,1N}(\omega)&\Gamma_{R,2N}(\omega)&\cdots&\Gamma_{N}(\omega)
 \end{pmatrix}
 \end{align}
is interpreted by a non-Hermitian Hamiltonian that describes the dissipatively-coupled magnons \cite{waveguide4,magnon_trap,magnon_trap_exp,Xiang}. Note that the coupling constant $\Gamma(\omega)$ still depends on the frequency.
 Then from Eq.~(\ref{eqn:T_matrix}), the 
 ${T}$-matrix becomes
 \begin{align}
 \begin{split}
 T_{k'k}&=\delta_{k'k}+\frac{1}{\omega_{k}-\omega_{k'}+i\eta}
 {\cal M}_{k'}^*
 G_N(\omega_{k})
 {\cal M}_{k}^T
 \end{split},
 \end{align}
 with ${\cal M}_{k}=\begin{pmatrix}g_{k,1}&\cdots&g_{k,N}
 \end{pmatrix}$.

 When propagating through magnetic nanowire arrays, the surface phonon of momentum $k$ is scattered between different states, and the transmitted waves (assuming $k>0$)
 \[
 \lim_{y\rightarrow +\infty}\psi_t(y)=\sum_{k'}\langle y|k'\rangle T_{k'k}\]
 and reflected waves
 \[\lim_{y\rightarrow -\infty}\psi_r(y)=\sum_{k'}\langle y|k'\rangle T_{k'k}
 \]
  at position far away from the magnetic wires are determined by the $T$-matrix.
Supposing an initial plane wave $e^{iky}$, with the transmitted and reflected waves $\psi_t(y)$ and $\psi_r(y)$, 
 the elements of the phonon scattering matrix are given by  \cite{scattering_PRB,scattering_PRE,Mahan}
 \begin{align}
 \nonumber
 &S_{21}(\omega_{k})=e^{ikD}
 \left(1-\frac{i}{c_r}
 {\cal M}_{k}^*
 G_N(\omega_{k})
 {\cal M}_{k}^T\right),\\
 &S_{11}(\omega_{k})=-\frac{i}{c_r}
 {\cal M}_{-k}^*
 G_N(\omega_{k}){\cal M}_{k}^T.
 \label{phonon_transmission}
 \end{align}
 where $D$ is the propagation length of the SAWs.
 The unitarity of the scattering matrix is guranteed by $|S_{21}(k)|^2+|S_{11}(k)|^2=1$ when $\alpha_G\rightarrow 0$. Nevertheless, the unitarity is broken when there exists magnon damping $\alpha_G$.
 
 The experiments \cite{Xu,DMI} are typically performed with one magnet. We thus first calculate the scattering matrix of phonon in the presence of a single magnetic wire in the following \cite{scattering_PRE,scattering_PRB,chiral_emitter,Gardiner_equivalence,Gardiner_chiral}. In this situation, the magnon Green function is given by 
  $G_{N=1}(\omega)=1/({\omega-\omega_{\rm K}-\Sigma(\omega)})$
  with magnon self-energy 
  $\Sigma(\omega)=\sum_{k'}|g_{k'}|^2/\left(\omega-\omega_{k'}+i\eta_{k'}\right)=-{i}(|g_k|^2+|g_{-k}|^2)/(2c_r)$. We obtain the scattering matrix,
  \begin{align}
  \nonumber
  &S_{21}(k)=e^{ikD}\left(1-\frac{i}{c_r}\frac{|g_{k}|^2}{\omega_{k}-\omega_{\rm K}+i\alpha_G\omega_{\rm K}-\Sigma(\omega_{k})}\right),\\
  &S_{11}(k)=-\frac{i}{c_r}\frac{g_{k}g_{-k}^*}{\omega_{k}-\omega_{\rm K}+i\alpha_G\omega_{\rm K}-\Sigma(\omega_{k})}.
  \end{align} 
  It is seen that when the magnon-phonon coupling is complete nonreciprocal, we always have $S_{11}(k)=0$ no matter $g_k=0$ or $g_{-k}=0$; there is no reflection. Nevertheless, the transmission of phonon with opposite momenta depends on the nonreciprocity. At the resonance with $\omega_k=\omega_{\rm K}$, 
  \begin{align}
  S_{21}(k)=\left\{\begin{matrix}
  e^{ikL},\\
  \xi e^{ikL},
  \end{matrix}\right.\quad
  \begin{matrix}
  {\rm when}~~g_{k}=0,\\
  ~{\rm when}~~g_{-k}=0,
  \end{matrix}
  \end{align}
  where 
  \begin{align}
  \xi=\frac{2\alpha_G\omega_{\rm K}-|g_k|^2/c_r}{2\alpha_G\omega_{\rm K}+|g_k|^2/c_r}
  \end{align}
  modulates the amplitude of the transmitted waves. Generally, $\xi<1$ as $\alpha_G>0$, implying a suppression of transmission. Nevertheless, a negative $\alpha_G$, i.e., a gain \cite{gain1,gain2,gain3,Peng_gain}, can amplify the phonons as $|\xi|>1$. When $g_k=0$, the nanomagnet is not excited at all and the propagating SAWs of momentum $k$ only accumulate a propagation phase $kD$. While when $g_{-k}=0$, the SAWs are first absorbed and then emitted by the nanomagnet, resulting in a double dissipative phase shift $\pi/2$. This can be observed by a wire with a high magnetic quality $2\alpha_G\omega_{\rm K}\ll |g_k|^2/c_r$, leading to $\xi=e^{i\pi}$, i.e., without amplitude suppression but a pure phase shift. When $2\alpha_G\omega_{\rm K}\rightarrow |g_k|^2/c_r$, the transmission tends to be zero and the energy accumulates in the magnet \cite{magnon_trap}. Finally, when $2\alpha_G\omega_{\rm K}\gg |g_k|^2/c_r$, $\xi\rightarrow 1$ and the modulation vanishes and the phonon diode effect is very small.
  
\begin{figure*}[tbh]
  	\begin{minipage}[]{18cm}
  		\parbox[t]{8cm}{
  			\includegraphics[width=7.85cm]{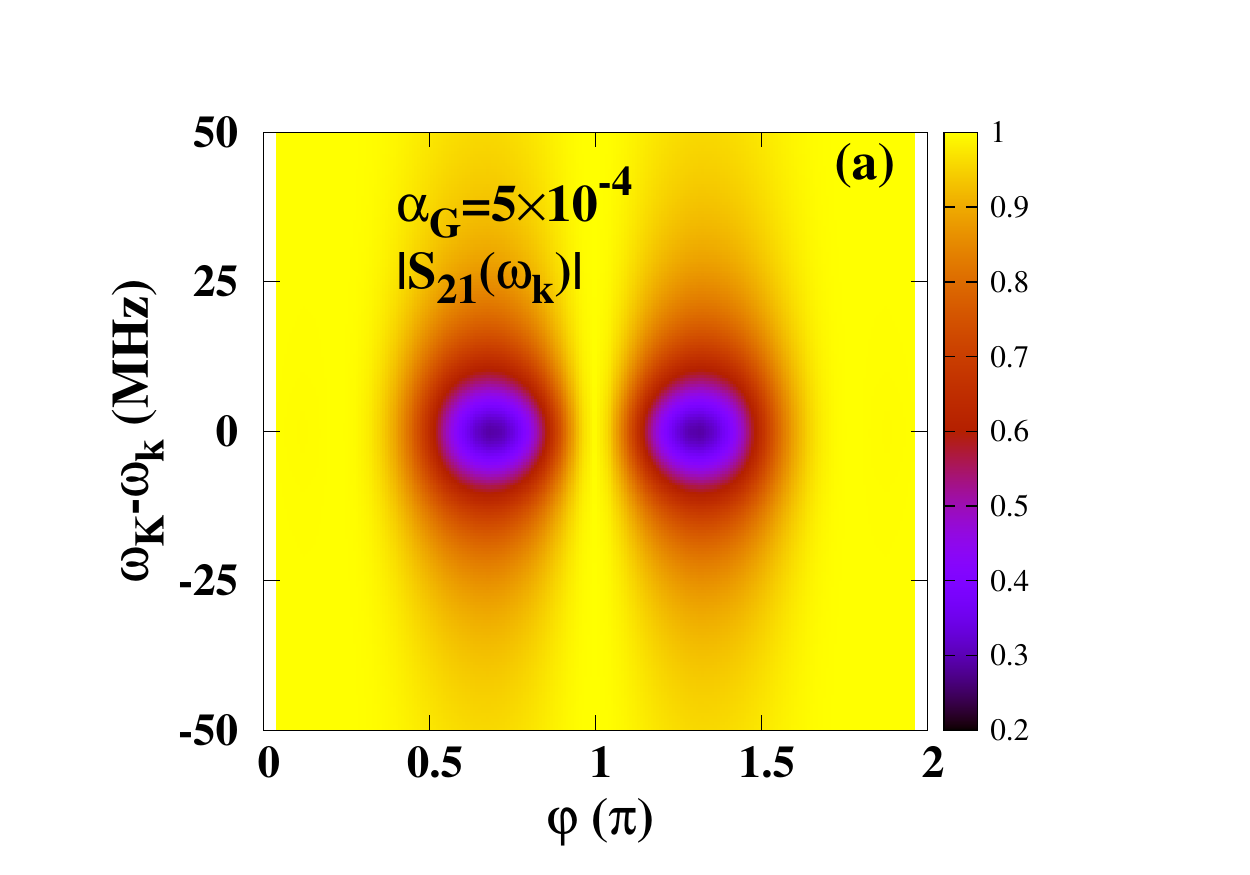}}
  		\parbox[t]{8cm}{
  			\includegraphics[width=7.85cm]{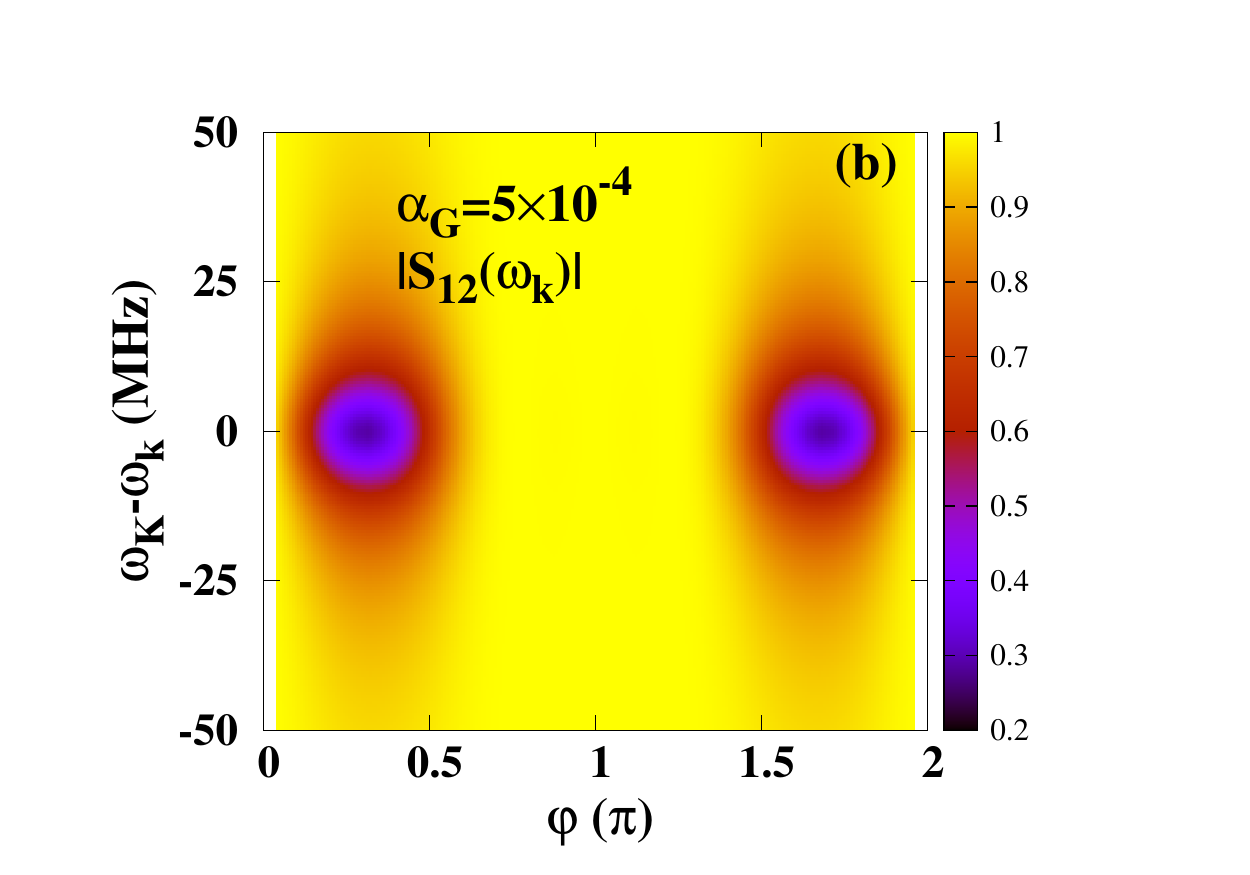}}
  		\vskip -0.6cm
  		\parbox[t]{8cm}{
  			\includegraphics[width=7.85cm]{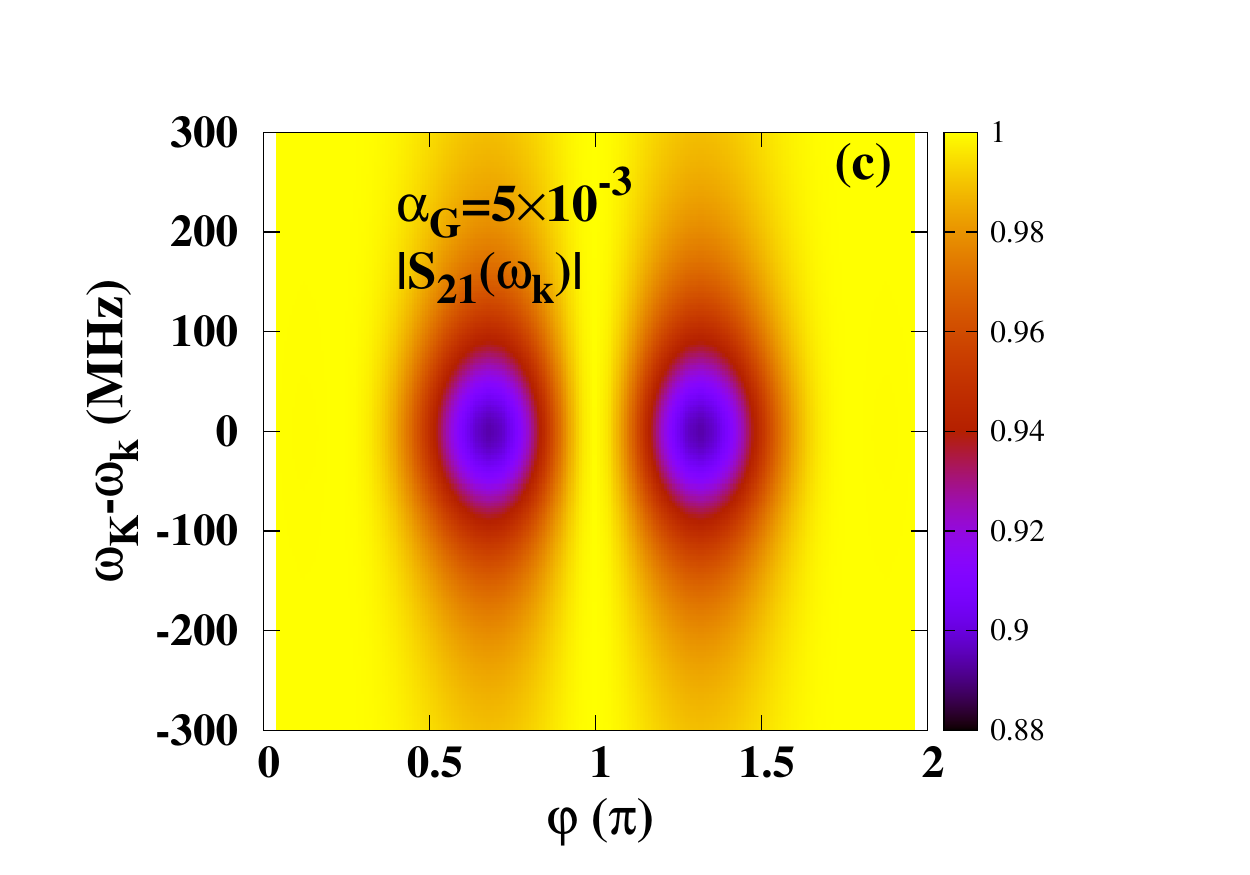}}
  		\parbox[t]{8cm}{
  			\includegraphics[width=7.85cm]{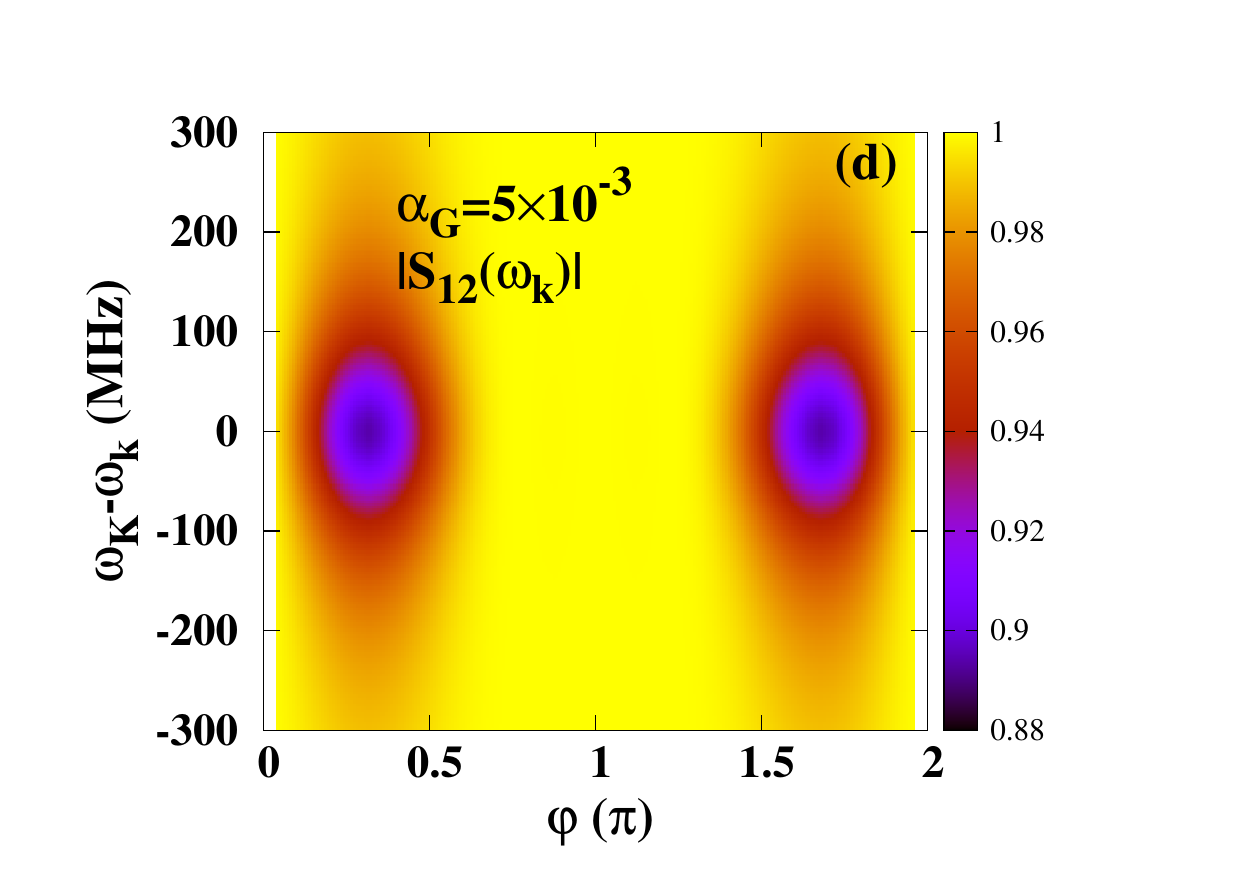}}
  	\end{minipage}
  	\begin{minipage}[]{16.2cm}
  		\begin{center}
  			\caption{(Color online)
  				Absolute value of
  				the phonon transmission $|S_{21}(\omega_k)|$ [(a) and (c)] and $|S_{12}(\omega_k)|$ [(b) and (d)] with respect to the field direction $\varphi$ and the magnon frequency $\omega_{\rm K}$ of one Ni nanowire, around the phonon frequency $\omega_k=2\pi\times 20$~GHz.  We take a small and large Gilbert dampings $\alpha_G=5\times 10^{-4}$ [(a) and (b)] and $5\times 10^{-3}$ [(c) and (d)], respectively.
  				The material parameters are given in the text.}
  			\label{fig:phonon_diode}
  		\end{center}
  	\end{minipage}
  \end{figure*}

In Fig.~\ref{fig:phonon_diode}, we plot the absolute value of
 the phonon transmission $|S_{21}(\omega_k)|$ and $|S_{12}(\omega_k)|$ with respect to the field direction $\varphi$ and the magnon frequency $\omega_{\rm K}$ of one (e.g, Ni) nanowire, around the phonon frequency $\omega_k=2\pi\times 20$~GHz. We take a small and large Gilbert dampings $\alpha_G=5\times 10^{-4}$ [(a) and (b)] and $5\times 10^{-3}$ [(c) and (d)], respectively, to illustrate effects of different magnetic qualities. The other material parameters are the same as those in Fig.~\ref{fig:phonon_chirality}. With both Gilbert dampings, the phonon transmission shows a dip around the critical angles (e.g., $\phi_c\approx 0.73 \pi$ and $1.27\pi$ with $g_{-|k|}=0$) and resonance frequency $\omega_{\rm K}=\omega_k$. Around these angles and
 frequencies, $|S_{12}(\omega)|\ne |S_{21}(\omega)|$, demonstrating the nonreciprocity and phonon diode effect \cite{Xu,DMI}. The energy range for the diode effect is broadened by a large Gilbert damping, but the magnitude is significantly suppressed, suggesting a need to improve the magnetic quality for a clear experimental observation and application when one magnet is employed. As the magnon damping breaks the unitarity of the phonon scattering matrix, the phonon suffers from the ``resistivity" when passing through the magnets. If this resistivity is enhanced by the number of magnets, we expect that the diode effect can be enhanced when there are many magnets, as studied in the following subsection.

  \subsection{Phonon resistivity by collective magnon modes}
  \label{collective_modes}
  
  Although the phonon is assumed to have a small damping here, the magnon damping can be large. The conversion of phonon to magnon then suffers from a large damping that brings a resistivity. Experimentally, this can correspond to a short propagation length for phonon that scales with the number of magnet. To this end, we generally calculate the phonon transmission through many (identical) magnetic wires.

  We express the phonon scattering matrix by the collective modes of magnons via the eigenvectors of the non-Hermitian Hamiltonian ${\cal H}_N(\omega_k)$ \cite{waveguide4}. Assuming the right eigenvectors of ${\cal H}_N(\omega_k)$ are $\psi_{\zeta}$ with corresponding eigenvalue $\nu_{\zeta}$, ${\cal H}_N(\omega_k)\psi_{\zeta}=\nu_{\zeta}\psi_{\zeta}$. Here, $\zeta=\{1,\cdots,N\}$ labels the collective modes. We also define the right eigenvectors $\phi_{\zeta}$ of ${\cal H}^{\dagger}_N(\omega_k)$ with corresponding eigenvalue $\nu_{\zeta}^*$. The eigenvectors satisfy the orthornormal conditions with $\psi^{\dagger}_{\zeta}\phi_{\zeta'}=\delta_{\zeta\zeta'}$ and $\phi_{\zeta}^{\dagger}\psi_{\zeta'}=\delta_{\zeta\zeta'}$ \cite{waveguide4}. Thus, the magnon (retarded) Green function is found to be
  \begin{align}
  G^r_N(\omega)=\sum_{\zeta}\psi_{\zeta}\phi_{\zeta}^{\dagger}\frac{1}{\omega-\nu_{\zeta}},
  \label{retarded}
  \end{align}
 which is determined by the collective modes of the wires. Therefore, the net effect of many wires is not a simple summation of that of single wire, which would bring a modulation factor $\xi^N$, but may lead to different features. The phonon transmission can thus demonstrate the existence and information of the collective mode of many magnetic wires. Many properties of the collective mode were addressed in our previous works \cite{waveguide4,magnon_trap}.

\begin{figure}[ht]
 	\centering
 	\includegraphics[width=7.1cm]{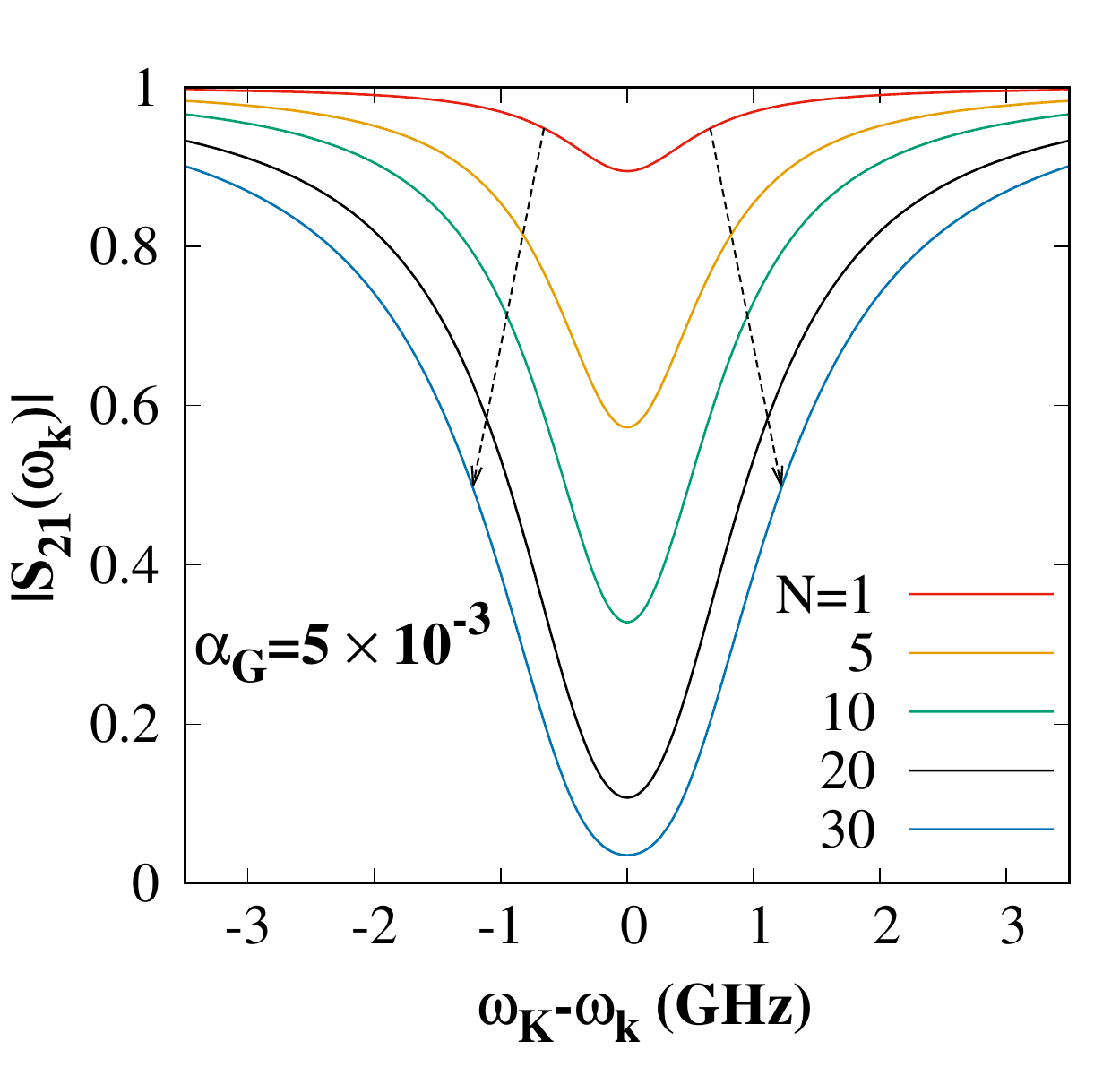}
 	\caption{Dependence on magnet number $N$ of the  phonon transmission $|S_{21}(\omega_k)|$ for Ni nanowire on GGG substrate. The damping of wire is chosen to be relatively large with $\alpha_G=5\times 10^{-3}$. The direction of the magnetic field is chosen to be at the critical angle such that $|S_{12}(\omega_k)|=1$. The dashed arrows indicate the evolution of the half width of the spectra with the increase of the wire number. The other material parameters are given in the text. }
 	\label{fig:phonon_resistivity}
 \end{figure}

 We numerically diagonalize the non-Hermitian Hamiltonian and calculate the phonon transmission through a magnetic array with distance $\delta$ between the neighbouring wires. Such an array was commonly employed to excite short-wavelength spin waves on top of a magnetic film \cite{magnon_magnon3,Jilei_Ni}. By taking $\delta=3.2\pi/k$ and the other parameters in the last subsection, we show the improvement of the diode effect by many magnetic wires. Figure~\ref{fig:phonon_resistivity} is a plot of the phonon transmission $|S_{21}(\omega_k)|$ at the critical angles with $|S_{12}(\omega_k)|=1$ when the damping of magnetic wire is large with $\alpha_G=5\times 10^{-3}$. Although the transmission is still very large with one magnetic wire, it tends to be zero rapidly with tens of wires. The frequency window of the filtering also increases with the increase of the wire number, i.e., a broadband nonreciprocity, as indicated by the dashed arrows in the figure, implying the advantage of many-wire configuration. The suppression of the phonon transmission is not as rapid as $\xi^N$-law with the increase of magnet number, indicating that it is the collective modes that play roles in the filtering. Note that the phonon reflection is zero in this complete nonreciprocal case, implying that the phonons are damped by exciting the magnon collective modes as the total number of magnon and phonon is conserved in the linear regime. 
 
 \section{Phase-sensitive microwave transmission}
 
 \label{sec:transmission}
 
 The phase-sensitive microwave scattering matrix is also an efficient way to detect the nonreciprocity \cite{magnon_magnon3,magnon_trap_exp}. We again formulate a general case with $N$ parallel magnetic wires. We consider the microwave excites the $i$-th wire and the radiated microwave is read out above the $j$-th wire. Such a setup was realized experimentally with two  wires by narrow striplines on top of the wires \cite{magnon_trap_exp}. Following our previous works \cite{magnon_magnon3,waveguide4,Xiang}, the equation of motion of the magnons augmented by the microwave input leads to \cite{Gardiner_chiral,Gardiner_equivalence}
 \begin{align}
 -i\omega\hat{B}(\omega)=-i\hat{H}_{N}(\omega)\hat{B}(\omega)-\sqrt{\kappa_p}\hat{P}_{\rm in}(\omega),
 \label{EOM}
 \end{align}
where $\hat{B}=(\hat{\beta}_1,\hat{\beta}_2,\cdots,\hat{\beta}_N)^T$ and $\kappa_p$ is the dissipative damping of magnon by microwave radiation, which we assume to be much smaller than the intrinsic Gilbert damping. $\hat{P}_{\rm in}$ is the input of microwave photon in which we assume the $i$-th wire is excited by the local active stripline and  the element is $(\hat{P}_{\rm in})_l=\delta_{li}\hat{p}_{\rm in}$. The local input microwaves, actually, excite all the magnetic wires as they are coupled through virtual exchange of the surface phonons. These excited wires can radiate out the microwaves that can be detected by the passive stripline. We monitor the microwave at the $j$-th wire with the photon output  \cite{Gardiner_chiral,Gardiner_equivalence,magnon_magnon4} 
\begin{align}
\hat{p}_{\rm out}(\omega)=\hat{p}_{\rm in}(\omega)\delta_{ij}+\sqrt{\kappa_p}\hat{\beta}_j(\omega).
\end{align} 
The magnon excitation at the $j$-th wire is represented by magnon Green function from Eq.~(\ref{EOM}),
\begin{align}
	\hat{\beta}_j(\omega)=i\sqrt{\kappa_p}(G_N(\omega))_{ji}\hat{p}_{\rm in}(\omega).
\end{align}
When $j=i$, we obtain the microwave reflection at the $i$-th wire,
\begin{align}
\nonumber
\tilde{S}_{ii}(\omega)&=\hat{p}_{\rm out}(\omega)/\hat{p}_{\rm in}(\omega)=1-i\kappa_p(G_N(\omega))_{ii}\\
&=1-i\kappa_p\sum_{\zeta=1}^N\frac{\left(\psi_{\zeta}(\omega)\right)_i\left(\phi_{\zeta}^{\dagger}(\omega)\right)_i}{\omega-\nu_{\zeta}},
\end{align}
which reads out the diagonal term of the magnon Green function. While when $j\ne i$, we obtain the microwave transmission from the $i$-th wire to the $j$-th one,
\begin{align}
\nonumber
\tilde{S}_{ji}(\omega)&=-i\kappa_p(G_N(\omega))_{ji}\\
&=-i\kappa_p\sum_{\zeta=1}^N\frac{\left(\psi_{\zeta}(\omega)\right)_j\left(\phi_{\zeta}^{\dagger}(\omega)\right)_i}{\omega-\nu_{\zeta}},
\end{align}
which is expressed by the off-diagonal term of the magnon Green function. Therefore, an ergodic detection of all the microwave reflection and transmission can give the whole magnon Green function, whose inverse gives all the terms of the magnon Hamiltonian that contains rich information.

 The simplest experimental setup employs two identical wires \cite{magnon_magnon4,magnon_trap_exp}, in which case the magnon Green function reads
 \begin{align}
 \nonumber
 G_2(\omega)&=\frac{1}{(\omega-\tilde{\omega}_{\rm K}+i\Gamma_1(\omega))^2+\Gamma_{R,12}(\omega)\Gamma_{L,21}(\omega)}\\
 &\times\left(\begin{matrix}
 \omega-\tilde{\omega}_{\rm K}+i\Gamma_1(\omega)&-i\Gamma_{L,21}(\omega)\\
 -i\Gamma_{R,12}(\omega)&\omega-\tilde{\omega}_{\rm K}+i\Gamma_1(\omega)
 \end{matrix}\right).
 \end{align}
Accordingly, the microwave reflection and transmission read
\begin{align}
\nonumber
&\tilde{S}_{11}(\omega)=1-\frac{i\kappa_p(\omega-\tilde{\omega}_{\rm K}+i\Gamma_1(\omega))}{(\omega-\tilde{\omega}_{\rm K}+i\Gamma_1(\omega))^2+\Gamma_{L,21}(\omega)\Gamma_{R,12}(\omega)},\\
&\tilde{S}_{21}(\omega)=-\frac{\kappa_p\Gamma_{R,12}(\omega)}{(\omega-\tilde{\omega}_{\rm K}+i\Gamma_1(\omega))^2+\Gamma_{L,21}(\omega)\Gamma_{R,12}(\omega)},
\end{align}
recovering our previous results \cite{magnon_magnon3,magnon_magnon4,magnon_trap_exp}. We are particularly interested in the resonant situation with $\omega=\omega_{\rm K}$. Recalling $\Gamma_{R,12}(\omega)=(|g_{k_{\ast}}|^2/c_r)e^{i\omega|y_2-y_1|/c_r}$, the real part of the microwave transmission with complete nonreciprocity $\Gamma_{L,21}=0$ reads
\begin{align}
{\rm Re}\left(\tilde{S}_{21}(\omega)\right)=\frac{2\kappa_p\Gamma_{1}(\omega)}{(\alpha_G\omega+\Gamma_1(\omega))^2}\sin\left(\frac{\omega|y_2-y_1|}{c_r}\right),
\end{align}
which oscillates with the microwave frequency. The frequency difference between two neighboring peak of ${\rm Re}\left(\tilde{S}_{21}(\omega)\right)$ is
\begin{align}
\Delta\omega=2\pi c_r/|y_2-y_1|,
\label{frequency_relation}
\end{align}
which is sensitive to the wire distance and phonon group velocity.

\begin{figure}[ht]
 	\centering
 	\includegraphics[width=7.4cm]{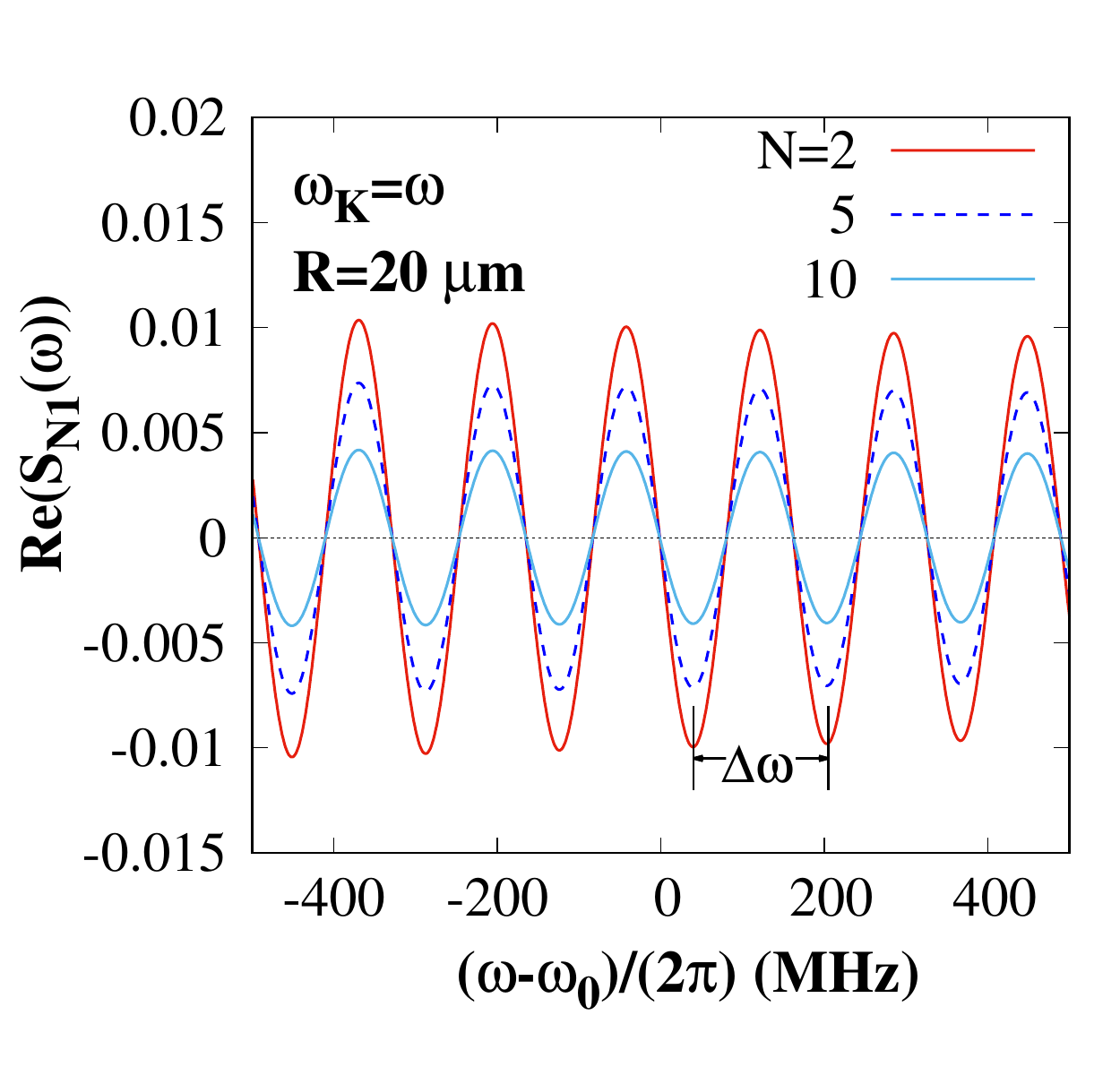}
 	\caption{Real part of the microwave transmission between the nanowires at the left and right edges at the resonant condition $\omega_{\rm K}=\omega$. We adopt different number of Ni nanowires $N=\{2,5,10\}$ on top of GGG and tune the magnetic field to the complete chirality with $\tilde{S}_{1N}(\omega)=0$. The damping of wire is chosen to be relatively large with $\alpha_G=5\times 10^{-3}$. The other material parameters are given in the text. }
 	\label{fig:microwave_transmission}
 \end{figure}

The microwave transmission with two-wire setup was studied in our previous works \cite{magnon_magnon3,magnon_magnon4,magnon_trap_exp}. Here, the Green function formalism allows us to study the influence of the middle wires on the microwave transmission.
We find relation Eq.~(\ref{frequency_relation}) is very robust even when the middle wires are added, while the amplitude of the microwave transmission is suppressed by the middle wires. We plot the real part of the transmission amplitude in Fig.~\ref{fig:microwave_transmission} as a
 function of microwave frequency close to $\omega_{0}=2\pi\times20~\mathrm{GHz}$. We tune the nanowire Kittel mode to be resonant to the microwave frequency. In the calculation, we take width $w=250~\mathrm{n m}$ and
 thickness $d=30~\mathrm{nm}$ for the Ni nanowires on top of GGG. The distance between the two wires at the left and right edges is $R=20~\mu$m. The intrinsic magnetic damping is
 chosen as $\kappa_{m}=5\times10^{-3}$ and the radiative damping $\kappa_{\omega
 }=2\pi\times 10~\mathrm{MHz}$. We observe the oscillation of the microwave transmission (real part) with respect to the microwave frequency, which is robust to the added middle wires. We note that the transmission does not vanish at the nodes as it becomes
purely imaginary. The calculated frequency difference between the neighboring peaks (dips) $\Delta\omega=2\pi \times 164$~MHz, agreeing well with Eq.~(\ref{frequency_relation}). 
 
The experiment \cite{phonon_long1} observed the oscillation of the microwave reflection with respect to the microwave frequency in the YIG|GGG|YIG sanwidched structure. There, the forth and back of the phonons between the two YIG layers are responsible. Here, the calculated oscillation in our transverse structure has a different physical origin. On one hand, the oscillation appears in the microwave transmission rather than reflection. On the other hand, the coupling is chiral and the excited phonon is unidirectional rather than a back and forth motion. 
 
 \section{Nonreciprocal phonon pumping}
 \label{sec:Landau_Buttiker}
 
 In the presence of the nonreciprocal magnetoelastic coupling, the injected phonon from a precessing magnetization that travels with group velocity $c_r$ forms opposite currents at the two sides of the nanowires, i.e., a chiral pumping effect that was considered recently in magnonics and photonics in the presence of one to many magnets \cite{magnon_magnon1,magnon_magnon2,magnon_magnon3,magnon_magnon4,waveguide3,waveguide4,magnon_electron}. In this part, we extend the formalism in terms of the Green function to calculate the chiral phonon pumping and make a connection to the phonon scattering matrix. 
 
 As the group velocity of surface phonon does not depend on the momentum, the current $I_{L/R}$ in the $\pm \hat{\bf y}$-direction is solely determined by the total injection rate ${\cal P}_{R/L}$ for the left- and right-moving phonons with
 	\begin{align}
 	I_{R/L}={\cal P}_{R/L}c_r.
 	\end{align} 
 The corresponding phonon injection rate depends on the nonequilibrium distributions and can be generally calculated by Landauer-B\"uttiker formula \cite{Abrikosov,Fetter,Mahan,Haug}. The injection rates read
 \begin{align}
 \nonumber
 {\cal P}_{R(L)}&=\frac{i}{\hbar}\left\langle \left[\hat{H},\sum_l\hat{\beta}_l^{\dagger}\hat{\beta}_l\right]\right\rangle\\
 &=-2{\rm Re}\left(\sum_l\sum_{k>0(<0)}g_{k,l}^*G^{<}_{lk}(t,t)\right),
 \end{align}
 which is evaluated by the Green functions
 \[G^<_{lk}(t,t')\equiv -i\left\langle \hat{b}_{k}^{\dagger}(t')\hat{\beta}_l(t)\right\rangle,\]
 and 
 \[G^<_{kl}(t,t')\equiv -i\left\langle \hat{\beta}_l(t')\hat{b}_{k}(t)\right\rangle.
 \]
 The ``lesser" Green function can be calculated by invoking the time-ordered Green function
 \[G^t_{lk}(t,t')=-i\theta(t-t')\langle \hat{\beta}_l(t)\hat{b}_{k}^{\dagger}(t')\rangle-i\theta(t'-t)\langle \hat{b}_{k}^{\dagger}(t')\hat{\beta}_l(t)\rangle,\]
 which evolves according to the coupled Hamiltonian $\hat{H}_c$:
 \begin{equation}
 \Big(-i\frac{\partial}{\partial t'}-\omega_{k}\Big)G^t_{lk}(t,t')=\sum_{l'} g_{k,l'}{G}_{ll'}^t(t,t'),
 \end{equation}
 where we define the time-ordered Green function for the Kittel magnons \[
 {G}^t_{ll'}(t,t')=-i\theta(t-t')\langle \hat{\beta}_l(t)\hat{\beta}_{l'}^{\dagger}(t')\rangle-i\theta(t'-t)\langle\hat{\beta}_{l'}^{\dagger}(t')\hat{\beta}_l(t)\rangle.
 \]
 Then through the operator $-i{\partial}/{\partial t'}-\omega_{k}\equiv [{\cal G}^t_{k}(t')]^{-1}$, we express  
 \begin{equation}
 G_{lk}^t(t,t')=\sum_{l'}\int dt_1 g_{k,l'}{G}_{ll'}^t(t,t_1){\cal G}_{k}^t(t_1,t').
 \end{equation}
 The time-contoured  Green function shares the same Feynman rule as the time-ordered one, which allows us to 
 use Langreth theorem \cite{Haug} to obtain the ``lesser" Green function (in the frequency space) 
 \begin{equation}
 G_{lk}^<(\omega)=\sum_{l'}g_{k,l'}\big[{G}_{ll'}^r(\omega){\cal G}^<_{k}(\omega)+{G}_{ll'}^<(\omega){\cal G}_{k}^a(\omega)\big].
 \end{equation}
 Thus, the pumping rate is calculated to be  
 \begin{align}
 \nonumber
 {\cal P}_{R(L)}&=-2\sum_{k>0(<0)}\sum_{ll'}\int \frac{d\omega}{2\pi}{\rm Re}({g_{k,l}^*G^r_{ll'}(\omega)}g_{k,l'}{\cal G}^<_{k}(\omega)\\
 &+g_{k,l}^*G^<_{ll'}(\omega)g_{k,l'}{\cal G}^a_{k}(\omega)).
 \label{eqn:photon_pumping}
 \end{align}
 With complete nonreciprocity, one of ${\cal P}_{R}$ and ${\cal P}_{L}$ vanishes, indicating that the pumped phonon flows unidirectionally in half of the elastic substrate. 
 
 For the phonon, the ``lesser'' and advanced Green functions are given by ${\cal G}_{k}^<(\omega)=2\pi if(\omega)\delta(\omega-\omega_{k})$, with $f(\omega)=1/(e^{\omega/(k_BT)}-1)$ being the Bose-Einstein distribution at temperature $T$, and ${\cal G}_{k}^a(\omega)={1}/({\omega-\omega_{k}-i\eta_k})$, respectively. The retarded Green function of magnons $G_N^r(\omega)$ is given by Eq.~(\ref{retarded}), while the advanced Green function $G_N^a(\omega)=G_N^{r*}(\omega)$. These two Green function defines the spectra function ${\mathcal A}(\omega)=i(G^r_N(\omega)-G_N^a(\omega))$, and, from fluctuation-dissipation theorem, we have  \cite{Haug}
 \begin{align}
 \nonumber
 G_{ll'}^<(\omega)&=iF(\omega){\mathcal A}_{ll'}(\omega)\\
 &=-F(\omega)\left(G_N^r(\omega)-G_N^{r*}(\omega)\right)_{ll'},
 \end{align}
 where $F(\omega)$ parameterizes the magnon nonequlibrium distribution. Accordingly, we can demonstrate ${\cal I}(\omega)=\sum_{ll'}g_{k,l}^*G_{ll'}^<(\omega)g_{k,l'}$ is purely imaginary as ${\cal I}^*(\omega)=-{\mathcal I}(\omega)$. The phonon injection rates then read
 \begin{align}
 \nonumber
 {\mathcal P}_{R(L)}&=\sum_{k>0(<0)}\sum_{ll'}{\rm Re}\left(ig_{k,l}^*G_{ll'}^r(\omega_k)g_{k,l'}\right)\\
 &\times (F(\omega_k)-f(\omega_k))\nonumber\\
 &-\sum_{k>0(<0)}\sum_{ll'}{\rm Re}\left(ig_{k,l}^*G_{ll'}^r(\omega_k)g_{k,l'}\right)f(\omega_k)\nonumber\\
 &-\sum_{k>0(<0)}\sum_{ll'}{\rm Re}\left(ig_{k,l}^*G_{ll'}^{r*}(\omega_k)g_{k,l'}\right)F(\omega_k).
 \label{Landuer_Buttiker}
 \end{align}
 At the thermal equilibrium, $F(\omega_k)=f(\omega_k)$ and ${\mathcal P}_{R(L)}$ vanishes. The injection rate is closely related to the phonon transmission as $\sum_{ll'}ig_{k,l}^*G_{ll'}^r(\omega_k)g_{k,l'}=c_r(1-S_{21}(\omega_k))$ from Eq.~(\ref{phonon_transmission}). 
 
\begin{figure}[ht]
 	\centering
 	\includegraphics[width=7.4cm]{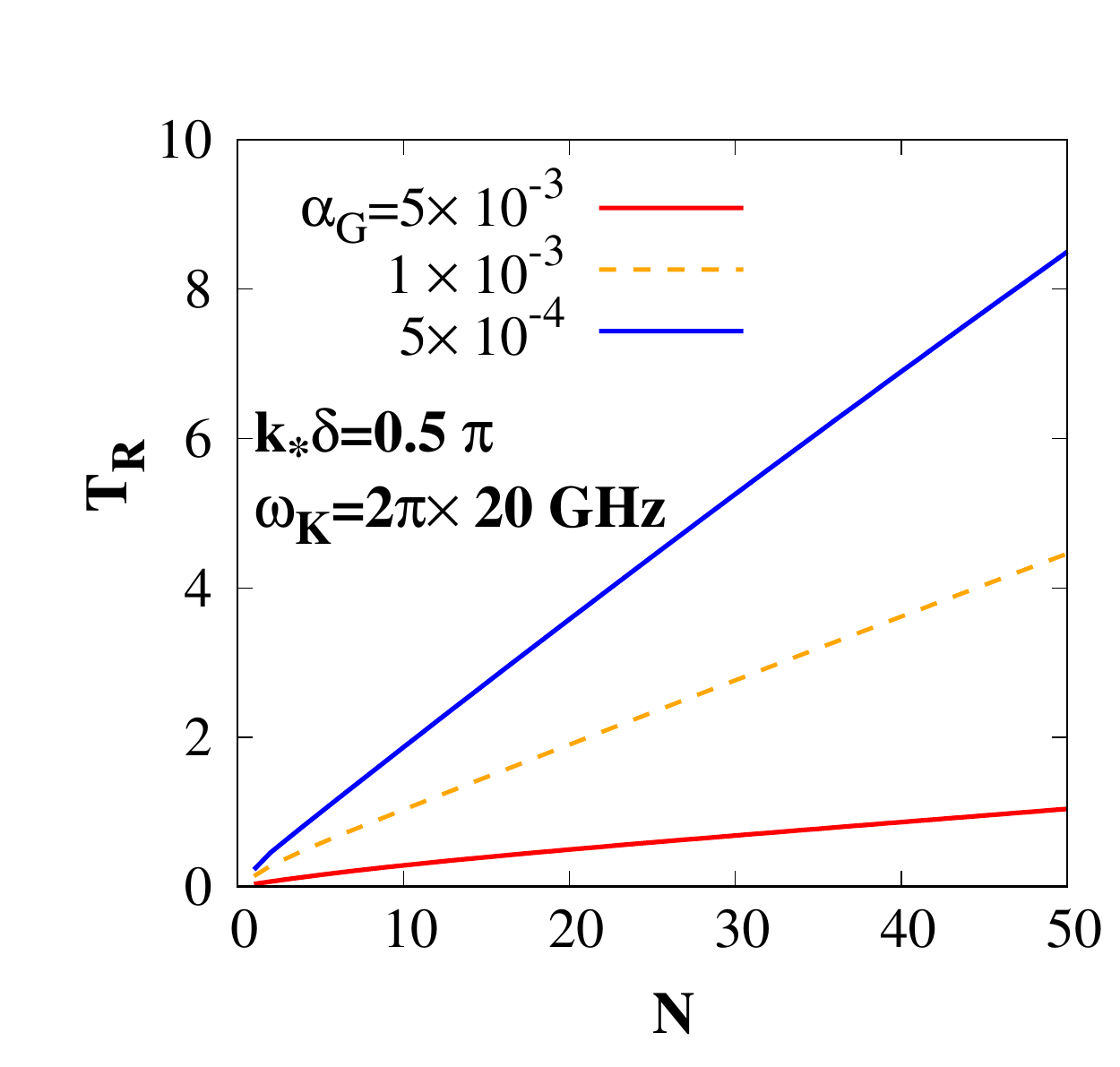}
 	\caption{Scaling of ${\cal T}_R$ with respect to the nanowire number with different magnetic qualities. We tune the magnetic field to the critical angle such that ${\mathcal T}_L=0$. The material parameters are indicated and given in the text. }
 	\label{fig:scaling}
 \end{figure}

 Usually, the injection rate with one magnetic nanowire is not large, and we may envision the injection rate can be improved with many nanowires.
 Therefore, we are particularly interested in the scaling relation with respect to the number of the nanowire.
To this end, we adopt, for simplicity, a monochromatic microwave of frequency $\omega_{\rm K}$ to resonantly excite the magnetic wires and $F(\omega_k)\rightarrow f(\omega_k)+\delta f\delta(\omega_k-\omega_{\rm K})$, with which Eq.~(\ref{Landuer_Buttiker}) is reduced to
 	\begin{align}
 	{\mathcal P}_{R(L)}=({1}/{c_r}){\mathcal T}_{R(L)}\delta f,
 	\end{align}
 	where 
 	\begin{align}
 	\nonumber
 	{\mathcal T}_R&=\frac{1}{2\pi}\sum_{ll'}{\rm Re}\left(ig_{k_*,l}^*(G_{ll'}^r(\omega_{\rm K})-G_{ll'}^{r*}(\omega_{\rm K}))g_{k_*,l'}\right),\\
 	{\mathcal T}_L&=\frac{1}{2\pi}\sum_{ll'}{\rm Re}\left(ig_{-k_*,l}^*(G_{ll'}^r(\omega_{\rm K})-G_{ll'}^{r*}(\omega_{\rm K}))g_{-k_*,l'}\right).
 	\end{align}
 	 Again, we employ the Ni nanowire on top of GGG with the same parameters as in Figs.~\ref{fig:phonon_resistivity} and \ref{fig:microwave_transmission} and tune the magnetic field to the critical angle such that ${\mathcal T}_L=0$.
 	 Figure~\ref{fig:scaling} shows the approximate linear scaling of ${\cal T}_R$ with respect to the nanowire number, suggesting an large injection rate with a nanowire array of high magnetic quality.  

 \section{Discussion and summary}
 \label{sec:summary}
 
 In conclusion, we formulate the nonreciprocal phonon transmission by acoustic pumping of magnon and its inverse process, the chiral pumping of phonon by ferromagnetic resonance. The model device we consider is one to many magnetic nanowires on top of a high-quality acoustic insulator rather than the extended magnetic film. We employ nanowire with thickness much smaller than the wavelength of the SAWs, in which situation the effect from the shear strain on the upper and lower surface cancels, different from that of thick films \cite{acoustic_pumping5,PRApplied}. Therefore, in the wire setup the nonreciprocity comes from the edge effect that is sensitive to the wire width and is strong when the wire width is comparable to the SAW wavelength. Both processes, associated with the phonon diode effect and unidirectional phonon current in half space, have high efficiency when the magnetic quality of the wire is high, but the efficiency is significantly enhanced by increasing the number of wire that allows to use material with ordinary magnetic quality to realize similar functionalities. The microwave transmission mediated by two remote magnetic wires that interact by virtual exchange of phonons is phase sensitive and hence can be used to detect, e.g., the phonon group velocity and wire distance, which could be a unique way to measure the coherent phonon propagation.
 
 The nonreciprocal magnon-phonon interaction comes from the chirality of magnon and rotation-momentum locking of surface phonon. Classically, we calculate the rotating forces at the edge of nanomagnet and show its relation with the magnon chirality. We use a quantum formalism and employ the Green function method to universally describe the chiral dynamics between magnon and other quasiparticles including photon \cite{waveguide3,waveguide4}, other magnon \cite{magnon_magnon1,magnon_magnon2,magnon_magnon3,magnon_magnon4}, electron \cite{magnon_electron} and phonon. We demonstrate the non-Hermitian interaction between wires lead to the collective motion of magnons that has influence on the phonon diode effect, microwave transmission, and chiral phonon pumping. 
 
 Magnetization dynamics can control the propagation of surface phonon in gigahertz frequency, much higher than the electric control in megahertz frequency \cite{megahertz1,megahertz2}. 
 Replacing the magnetic nanowire by the various nanomagnet configurations such as the nano-disks is an interesting extension. As addressed in the text, the gain by a negative Gilbert damping \cite{gain1,gain2,gain3,Peng_gain} can add amplification functionality on the basis of the nonreciprocity that could be useful in the future applications in the logic device.
 An inserted heavy metal between the nanomagnet and acoustic insulator may induce the Dzyaloshinskii-Moriya interaction that influences the nonreciprocity as well, which could be a possiblity to improve the magnetoelasitic nonreciprocity \cite{DMI,DMI2,DMI3}.

\begin{acknowledgments}
	This work was funded through the Emmy Noether Program of Deutsche Forschungsgemeinschaft (SE 2558/2-1).
	We thank Kei Yamamoto for sharing unpublished results 
	and bringing our attention to the recent experiments. We also thank Gerrit E. W. Bauer, Hanchen Wang and Xiang Zhang for useful discussions. 
\end{acknowledgments}

\appendix
	\section{Hamiltonian $\hat{H}_e$ and $\hat{H}_m$}
	\label{appendix}
	Here we address the Hamiltonian $\hat{H}_e$ and $\hat{H}_m$ used in the main text.
	 From the equation of motion, the SAW eigenmodes propagating in the $\hat{\bf y}$-direction of an isotropic elastic half space $(x<0)$ read \cite{SAW_book1,SAW_book2,Xiang}
	\begin{align}
	\nonumber
	\psi_{x}  &  =q\varphi_{k}\left(e^{qx}-\frac{2k^{2}}{k^{2}+s^{2}}e^{sx}\right)
	e^{iky},\\
	\psi_{y}  &  =ik\varphi_{k}\left(e^{qx}-\frac{2qs}{k^{2}+s^{2}}e^{sx}\right)
	e^{iky},
	\label{eqn:SAW_profile}
	\end{align}
	where $q=\sqrt{k^{2}-k_{l}^{2}}$ and $s=\sqrt{k^{2}-k_{t}^{2}}$ with
	$k_{l}=\omega_{k}\sqrt{\rho/(\lambda+2\mu)}$ and $k_{t}=\omega_{k}\sqrt{\rho/\mu}$ being the wave vectors for longitudinal and transverse bulk waves, respectively. Here, $\mu$ and $\lambda$ are the elastic Lam\'{e} constants, $\omega_{k}=c_{r}|k|$ represents the eigenfrequency of Rayleigh SAWs with velocity $c_{r}$, and $\varphi_{k}$ is a normalization constant. The opposite relative phase of the displacement field ${\mathrm{Arg}}(\psi_{y}/\psi_{x})|_{x=0}=\pm
	i$ for left- and right-propagating waves indicates the  rotation-momentum locking of SAWs.
	The displacement field $(\hat{u}_{x},\,\hat{u}_{y})$ is quantized by the eigenmodes $\boldsymbol{\psi}(k)$ and phonon operators $\hat{b}_{k}(t)$
	\begin{equation}
	\hat{\mathbf{u}}(x,y,t)=\sum_{k}\left[\boldsymbol{\psi}(x,y,k)\hat{b}_{k}(t)+\boldsymbol{\psi}^{\ast}(x,y,k)\hat{b}_{k}^{\dagger}(t)\right].
	\label{eqn:phonon_operator}
	\end{equation}
	The mode amplitudes $\boldsymbol{\psi}$ are then normalized to recover the elastic
	Hamiltonian of Rayleigh SAWs with
	\begin{equation}
	\hat{H}_{\mathrm{e}}=\rho\int d\boldsymbol{r}\,\dot{\hat{\boldsymbol{u}}}^{2}(x,z,t)=\sum_{k}\hbar\omega_{k}\hat{b}_{k}^{\dagger}\hat{b}_{k},
	\label{eqn:Hamiltonian_E}
	\end{equation}
	leading to the normalization condition
	\begin{equation}
	\int_{-\infty}^{0}dx\left(  |\psi_{x}|^{2}+|\psi_{y}|^{2}\right)  =\frac{\hbar}{2\rho L\omega_{k}}.
	\end{equation}
	We then obtain the normalization factor 
	\begin{equation}
	\varphi_k=\frac{1}{|k|}\frac{1+b^2}{2a(1-b^2)}\sqrt{\frac{2\hbar}{\rho L c_r}}\,\xi_P,
	\label{eqn:SAW_normalization}
	\end{equation}
	where the factor 
	\begin{equation}
	\xi_{\rm P}=\frac{a(1-b^2)}{1+b^2}\left(\frac{1+a^2}{2a}+\frac{2a(a-2b)}{b(1+b^2)}\right)^{-1/2},
	\label{eqn:ksi_P}
	\end{equation}
	with dimensionless material constants 
	\begin{equation}%
	\begin{split}
	a  &  =q/\left\vert k\right\vert =\sqrt{1-(c_{r}/c_{l})^{2}},\\
	b  &  =s/\left\vert k\right\vert =\sqrt{1-\eta^{2}}.
	\end{split}
	\label{eqn:alpha_beta}%
	\end{equation}
	Here $c_{r}=\eta\sqrt{\mu/\rho}$ and $c_{l}=\sqrt{(\lambda+2\mu)/\rho}$ are, respectively,  the sound velocities of the surface and longitudinal bulk waves.

\begin{figure}[th]
		\begin{center}
			{\includegraphics[width=8.5cm]{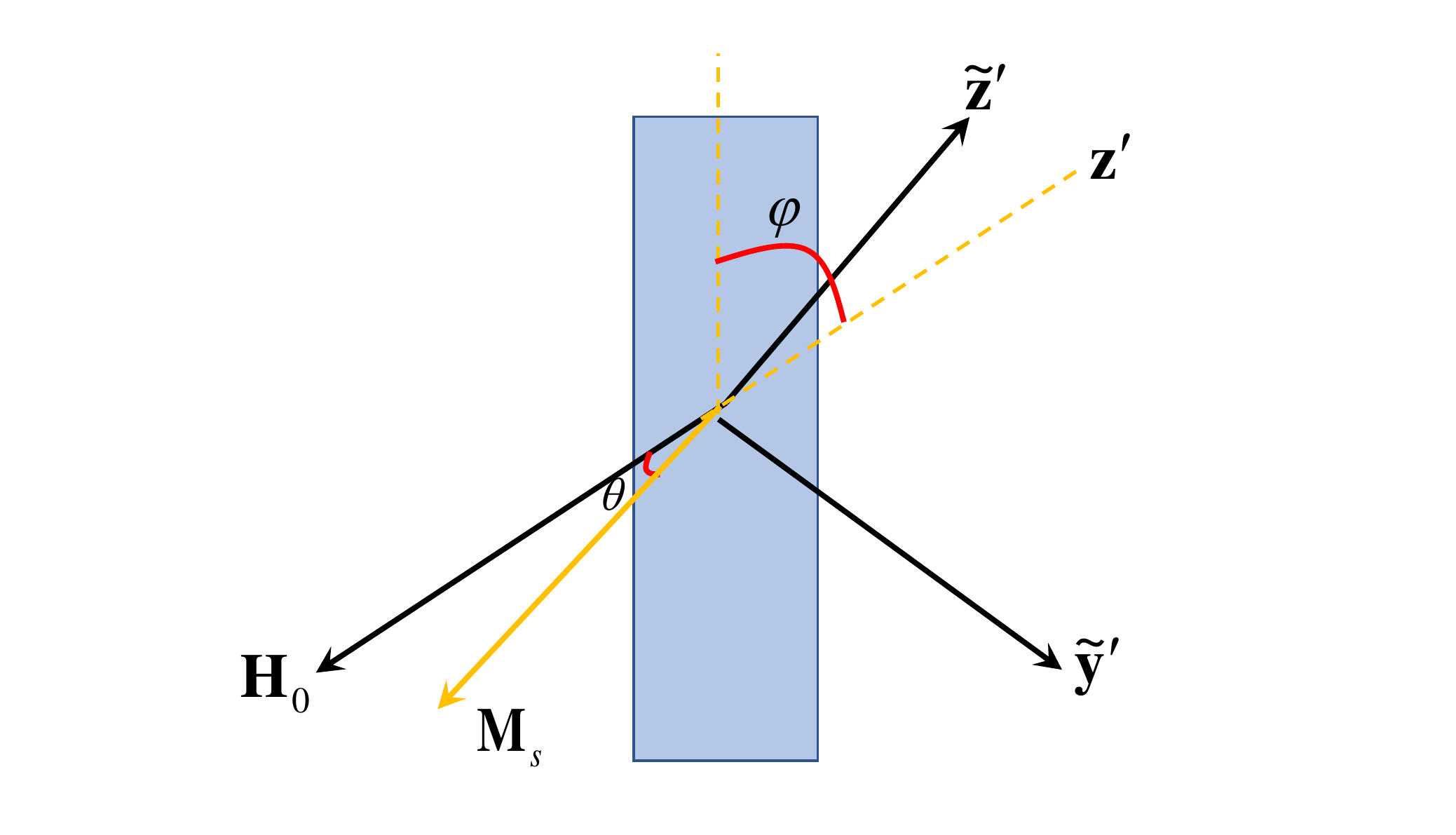}}
		\end{center}
		\caption{Configuration of equilibrium magnetization under an applied magnetic field of general direction.}%
		\label{configuration}%
	\end{figure}

	For the magnetic nanowire, we focus on the case with a large applied magnetic field $H_0$ to saturate the magnetization along the $z'$-axis (see Fig.~\ref{fig:phonon}). In the wire $\{xyz\}$-coordinate, $m_x=m_{x'}$, $m_y=m_{y'}\cos\varphi+m_{z'}\sin\varphi$ and $m_z=-m_{y'}\sin\varphi+m_{z'}\cos\varphi$.
	Focusing on the Kittel magnon, 
	\begin{equation}
	\hat{H}_{\mathrm{m}}=\mu_0\int d\boldsymbol{r}\,\left(  \hat{m}_{z'}H_{0}+\frac{1}{2}N_{xx}\hat{m}_x^{2}+\frac{1}{2}N_{yy}\hat{m}_y^{2}\right),
	\label{eqn:Hamiltonian_M}%
	\end{equation}
	where $\mu_{0}$ is the vacuum permeability, and $N_{xx}\simeq w/(d+w)$ and $N_{yy}\simeq d/(d+w)$ are the demagnetization constants with the nanowire width $w$ and thickness $d$ \cite{magnon_magnon2,magnon_trap}. Here the demagnetization factors are treated to be uniform across the wire by disregarding their spatial variation at the edges of nanowires.
	With the magnetic field applied opposite to the $\hat{\bf z}'$-direction (Fig.~\ref{fig:phonon} in the main text), the magnetization is not parallel to the applied magnetic field due to the demagnetization field. We thus assume that the magnetization is opposite to the $\tilde{\bf z}'$-direction, with an angle $\theta$ with respect to the $-\hat{\bf z}'$-direction, as shown in Fig.~\ref{configuration}. 
	The free-energy density in the static situation reads
	\begin{align}
	f_m=-\mu_0M_sH_0\cos\theta+\frac{1}{2}\mu_0N_{yy}(M_s\sin(\varphi-\theta))^2.
	\end{align}
	The minimum of the free energy $df_m/d\theta=0$ leads to 
	\begin{align}
	\theta=N_{yy}\sin\varphi\cos\varphi M_s/H_0,
	\label{theta}
	\end{align}
	when $H_0\gg M_s$, which is the situation we focus on.
	With knowing the static configuration, the effective magnetic field in the dynamical situation reads
	\begin{align}
	\nonumber
	{\bf H}_{\rm eff}&=-H_0\cos\theta\tilde{\bf z}'-H_0\sin\theta\tilde{\bf y}'-N_{xx}m_{x'}\tilde{\bf x}'\\
	\nonumber
	&-N_{yy}(m_{y'}\cos\varphi'+m_{z'}\sin\varphi')(\tilde{\bf y}'\cos\varphi'+\tilde{\bf z}'\sin\varphi')\\
	&\nonumber
	\approx [-H_0-N_{yy}(m_{y'}\sin\varphi\cos\varphi-M_s\sin^2\varphi)]\tilde{\bf z}'\\
	&-N_{xx}m_{x'}\tilde{\bf x}'-N_{yy}m_{y'}\cos^2\varphi\tilde{\bf y}',
	\end{align}
	in which $\varphi'=\varphi-\theta$ and we have used Eq.~(\ref{theta}) in the second step.
	With this magnetic field, the Landau-Lifshitz equation $d{\bf m}/dt=-\mu_0\gamma{\bf m}\times {\bf H}_{\rm eff}$ leads to
	\begin{align}
	\nonumber
	\frac{dm_{x'}}{dt}&=-\mu_0\gamma(-H_0-M_sN_{yy}\cos^2\varphi+N_{yy}M_s\sin^2\varphi)m_{y'},\\
	\frac{dm_{y'}}{dt}&=-\mu_0\gamma(H_0-N_{xx}M_s-N_{yy}M_s\sin^2\varphi)m_{x'},
	\end{align}
	from which we can find the Kittel frequency
	\begin{align}
	\nonumber
	\omega_{\rm K}&=\mu_0\gamma\sqrt{(H_0+M_sN_{yy}\cos(2\varphi))}\\
	\nonumber
	&\times \sqrt{(H_0+N_{xx}M_s-N_{yy}M_s\sin^2\varphi)}\\
	&\approx \mu_0\gamma H_0,
	\end{align}
	which generally exists when $H_0\gg M_s$.
	
	Our simple nanowire configuration is different from surface-modulated magnonic crystals or ferromagnetic nanogratings, where the Kittel mode is reported to exist in a region of magnetization direction with $\varphi \lesssim \pi/3$ \cite{referee1,referee2,referee3}. In the latter cases, the demagnetization field could develop a more complicated distribution due to the special geomerty of magnetizations.
	
	 The transverse magnetization is then quantized by the Kittel-magnon operator
	$\hat{\beta}(t)$ with wave function $\tilde{m}_{\chi=\{x',y'\}}({\bf r})$:
	\begin{equation}
	\hat{\mathbf{m}}_{\chi}({\bf r})=-\sqrt{2\gamma\hbar M_{s}}\left(  \tilde{m}_{\chi}({\bf r})\hat{\beta}(t)+\tilde{m}_{\chi}^{\ast}({\bf r})\hat{\beta}^{\dagger}(t)\right).
	\label{eqn:magnon_operator}
	\end{equation}
	The magnon amplitudes $\tilde{m}_{x',y'}$ satisfy the normalization condition \cite{magnon_magnon1}
	\begin{equation}
	\int d\mathbf{r}\left(\tilde{m}_{x'}({\bf r})\tilde{m}_{y'}^{\ast}({\bf r})-\tilde{m}_{x'}^{\ast}({\bf r})\tilde{m}_{y'}({\bf r})\right)=-\frac{i}{2}.
	\label{eqn:normalization_Kittel}
	\end{equation}
	With a large $H_0\gg M_s$, the magnon is circularly polarized with $\tilde{m}_{y'}=i\tilde{m}_{x'}$ such that   
	\begin{equation}
	\tilde{m}_{x'}\simeq \frac{1}{2\sqrt{Lwd}},\quad \tilde{m}_{y'}\simeq \frac{i}{2\sqrt{Lwd}},
	\end{equation}
	and $\hat{H}_{\mathrm{m}}=\hbar\omega_{\mathrm{K}}\hat{\beta}^{\dagger
	}\hat{\beta}$ with frequency $\omega_{\rm K}\simeq \mu_{0}\gamma H_0$.


\begin{thebibliography}{67}%
\makeatletter
\providecommand \@ifxundefined [1]{%
 \@ifx{#1\undefined}
}%
\providecommand \@ifnum [1]{%
 \ifnum #1\expandafter \@firstoftwo
 \else \expandafter \@secondoftwo
 \fi
}%
\providecommand \@ifx [1]{%
 \ifx #1\expandafter \@firstoftwo
 \else \expandafter \@secondoftwo
 \fi
}%
\providecommand \natexlab [1]{#1}%
\providecommand \enquote  [1]{``#1''}%
\providecommand \bibnamefont  [1]{#1}%
\providecommand \bibfnamefont [1]{#1}%
\providecommand \citenamefont [1]{#1}%
\providecommand \href@noop [0]{\@secondoftwo}%
\providecommand \href [0]{\begingroup \@sanitize@url \@href}%
\providecommand \@href[1]{\@@startlink{#1}\@@href}%
\providecommand \@@href[1]{\endgroup#1\@@endlink}%
\providecommand \@sanitize@url [0]{\catcode `\\12\catcode `\$12\catcode
  `\&12\catcode `\#12\catcode `\^12\catcode `\_12\catcode `\%12\relax}%
\providecommand \@@startlink[1]{}%
\providecommand \@@endlink[0]{}%
\providecommand \url  [0]{\begingroup\@sanitize@url \@url }%
\providecommand \@url [1]{\endgroup\@href {#1}{\urlprefix }}%
\providecommand \urlprefix  [0]{URL }%
\providecommand \Eprint [0]{\href }%
\providecommand \doibase [0]{http://dx.doi.org/}%
\providecommand \selectlanguage [0]{\@gobble}%
\providecommand \bibinfo  [0]{\@secondoftwo}%
\providecommand \bibfield  [0]{\@secondoftwo}%
\providecommand \translation [1]{[#1]}%
\providecommand \BibitemOpen [0]{}%
\providecommand \bibitemStop [0]{}%
\providecommand \bibitemNoStop [0]{.\EOS\space}%
\providecommand \EOS [0]{\spacefactor3000\relax}%
\providecommand \BibitemShut  [1]{\csname bibitem#1\endcsname}%
\let\auto@bib@innerbib\@empty


\bibitem {spintronics}S. D. Bader and S. S. P. Parkin, Annu. Rev. Condens.
Matter Phys. \textbf{1}, 71 (2010).

\bibitem {spintronics_RMP}I. \v Zuti\'c, J. Fabian, and S. Das Sarma, Rev.
Mod. Phys. \textbf{76}, 323 (2004).	
	

\bibitem {Marie}G. Wang, B. L. Liu, A. Balocchi, P. Renucci, C. R. Zhu, T.
Amand, C. Fontaine, and X. Marie, Nat. Commun. \textbf{4}, 2372 (2013).

\bibitem {graphene_diffusion}J. Ingla-Ayn\'es, M. H. D. Guimar\~aes, R. J.
Meijerink, P. J. Zomer, and B. J. van Wees, Phys. Rev. B \textbf{92},
201410(R) (2015).

\bibitem{graphene_RMP} A. Avsar, H. Ochoa, F. Guinea, B. \"Ozyilmaz, B. J. van Wees, and I. J. Vera-Marun, Rev. Mod. Phys. \textbf{92}, 021003 (2020).

\bibitem{YIG_centimeter} A. A. Serga, A. V. Chumak, and B. Hillebrands, J. Phys. D \textbf{43},
264002 (2010).
	
\bibitem{magnonics1} B. Lenk, H. Ulrichs, F. Garbs, and M. Muenzenberg, Phys. Rep. {\bf 507}, 107 (2011).
\bibitem{magnonics2} A. V. Chumak, V. I. Vasyuchka, A. A. Serga, and B. Hillebrands, Nat. Phys. {\bf 11}, 453 (2015).
\bibitem{magnonics3} D. Grundler, Phys. Rep. {\bf 11}, 407 (2016).
\bibitem{magnonics4}  V. E. Demidov, S. Urazhdin, G. de Loubens, O. Klein, V. Cros,
A. Anane, and S. O. Demokritov, Phys. Rep. {\bf 673}, 1 (2017).	

\bibitem{Simon} S. Streib, H. Keshtgar, and G. E. W. Bauer, Phys. Rev. Lett. \textbf{121}, 027202 (2018).

\bibitem{Kruglyak} O. S. Latcham, Y. I. Gusieva, A. V. Shytov, O. Y. Gorobets, and V. V. Kruglyak, Appl. Phys. Lett. \textbf{115}, 082403 (2019).

\bibitem{phonon_long1} K. An, A. N. Litvinenko, R. Kohno, A. A. Fuad, V. V.
Naletov, L. Vila, U. Ebels, G. de Loubens, H. Hurdequint,
N. Beaulieu, J. Ben Youssef, N. Vukadinovic, G.
E. W. Bauer, A. N. Slavin, V. S. Tiberkevich, and O.
Klein, Phys. Rev. B \textbf{101}, 060407(R) (2020).
\bibitem{phonon_long2} A. R\"uckriegel and R. A. Duine, Phys. Rev. Lett. \textbf{124},
117201 (2020).

\bibitem{acoustic_pumping1} M. Weiler, L. Dreher, C. Heeg, H. Huebl, R. Gross, M. S. Brandt, and S. T. B. Goennenwein, Phys. Rev. Lett. \textbf{106}, 117601 (2011).
\bibitem{acoustic_pumping2} M. Weiler, H. Huebl, F. S. Goerg, F. D. Czeschka, R. Gross, and S. T. B. Goennenwein, Phys. Rev. Lett. \textbf{108}, 176601 (2012).
\bibitem{acoustic_pumping3} L. Dreher, M. Weiler, M. Pernpeintner, H. Huebl, R. Gross, M. S. Brandt, and S. T. B. Goennenwein, Phys. Rev. B \textbf{86}, 134415 (2012).
\bibitem{acoustic_pumping4} Y. Yahagi, B. Harteneck, S. Cabrini, and H. Schmidt, Phys. Rev. B \textbf{90}, 140405(R) (2014).
\bibitem{acoustic_pumping5} R. Sasaki, Y. Nii, Y. Iguchi, and Y. Onose, Phys. Rev. B \textbf{95}, 020407(R) (2017).
\bibitem{acoustic_pumping6} P. Delsing, A. N. Cleland, M. J. Schuetz, J. Kn\"orzer, G. Giedke, J. I. Cirac, and {\it et al}., J. Phys. D: Appl. Phys. \textbf{52}, 353001 (2019).
\bibitem{acoustic_pumping7} J. Puebla, M. Xu, B. Rana, K. Yamamoto, S. Maekawa, Y. Otani, J. Phys. D: Appl.
Phys. \textbf{53}, 26 (2020).


\bibitem{acoustic_millimeter} B. Casals, N. Statuto, M. Foerster, A. H.-M\'inguez, R. Cichelero, P. Manshausen, A. Mandziak, L. Aballe, J. M. Hern'andez, and F. Maci\'a,
Phys. Rev. Lett. \textbf{124}, 137202 (2020).

\bibitem{SAW_book1} E. A. Ash, A. A. Oliner, G. W. Farnell, H. M. Gerard,
A. J. Slobodnik, and H. I. Smith, in {\it Acoustic Surface
	Waves} (Topics in Applied Physics) (Springer, Berlin,
2014).


\bibitem{Xu} M. R. Xu, K. Yamamoto, J. Puebla, K. Baumgaertl,
B. Rana, K. Miura, H. Takahashi, D. Grundler, S.
Maekawa, and Y. Otani, Sci. Adv. \textbf{6}, eabb1724 (2020).
\bibitem{DMI} M. K\"u$\ss${}, M. Heigl, L. Flacke, Andreas H\"orner, M. Weiler, M. Albrecht, and A. Wixforth, arXiv:2004.03535.
\bibitem{paper1976} P. R. Emtage, Phys. Rev. B \textbf{13}, 3063 (1976).



\bibitem{Xiang} X. Zhang, G. E. W. Bauer, and T. Yu, Phys. Rev. Lett. \textbf{125}, 077203 (2020).
 
\bibitem{magnon_magnon1} Y. Au, E. Ahmad, O. Dmytriiev, M. Dvornik, T. Davison, and V. V. Kruglyak, Appl. Phys. Lett. \textbf{100}, 182404 (2012). 

\bibitem{magnon_magnon2} T. Yu, C. P. Liu, H. M. Yu, Y. M. Blanter, and G. E. W. Bauer, Phys. Rev. B \textbf{99}, 134424 (2019).

\bibitem{magnon_magnon3} J. L. Chen, T. Yu, C. P. Liu, T. Liu, M. Madami, K. Shen, J. Y. Zhang, S. Tu, M. S. Alam, K. Xia, M. Z. Wu, G. Gubbiotti, Y. M. Blanter, G. E. W. Bauer, and H. M. Yu, Phys. Rev. B \textbf{100}, 104427 (2019).

\bibitem{magnon_magnon4} T. Yu, Y. M. Blanter, and G. E. W. Bauer, Phys. Rev. Lett. \textbf{123}, 247202 (2019).

 
 
 \bibitem{magnon_trap_exp} H. C. Wang, J. L. Chen, T. Yu, C. P. Liu, C. Y. Guo, H.
 Jia, S. Liu, K. Shen, T. Liu, J. Y. Zhang, M. A. Cabero Z,
 Q. M Song, S. Tu, L. Flacke, M. Althammer, M. Weiler,
 M. Z. Wu, X. F. Han, K. Xia, D. P. Yu, G. E. W. Bauer,
 and H. M. Yu, arXiv:2005.10452.






\bibitem{waveguide1} A. G. Gurevich, Radiotekh. Elektron. (Moscow) \textbf{8}, 780 (1963).
\bibitem{waveguide2} A. G. Gurevich and G. A. Melkov, \textit{Magnetization Oscillations
	and Waves} (CRC, New York, 1996).
\bibitem{waveguide3} L. Martin, U.S. Patent No. US3426297A (1966).
\bibitem{waveguide4} T. Yu, Y.-X. Zhang, S. Sharma, X. Zhang, Y. M. Blanter, and G. E. W. Bauer, Phys. Rev. Lett. \textbf{124}, 107202 (2020).

\bibitem{Tang} N. Zhu, X. Han, C.-L. Zou, M. R. Xu, and
H. X. Tang, Phys. Rev. A \textbf{101}, 043842 (2020).

\bibitem{cavity} W. C. Yu, T. Yu, and G. E. W. Bauer, Phys. Rev. B \textbf{102}, 064416 (2020).

\bibitem{antenna1} T. Schneider, A. A. Serga, T. Neumann, B. Hillebrands,
and M. P. Kostylev, Phys. Rev. B \textbf{77}, 214411 (2008).

\bibitem{antenna2} V. E. Demidov, M. P. Kostylev, K. Rott, P. Krzysteczko,
G. Reiss, and S. O. Demokritov, Appl. Phys. Lett. \textbf{95}, 2509 (2009).
\bibitem{antenna3} I. Bertelli, J. J. Carmiggelt, T. Yu, B. G. Simon, C. C. Pothoven, G. E. W. Bauer, Y. M. Blanter, J. Aarts, and T. van der Sar, arXiv:2004.07746.


\bibitem{magnon_electron} T. Yu and G. E. W. Bauer, Phys. Rev. Lett. \textbf{124}, 236801 (2020).

\bibitem{skin0} V. M. Martinez Alvarez, J. E. Barrios Vargas, and L. E. F. Foa Torres, Phys. Rev. B \textbf{97}, 121401(R) (2018).
\bibitem{skin1} S. Y. Yao and Z. Wang, Phys. Rev. Lett. \textbf{121}, 086803
(2018).
\bibitem{skin2} Z. Gong, Y. Ashida, K. Kawabata, K. Takasan, S. Higashikawa, and M. Ueda, Phys. Rev. X \textbf{8}, 031079 (2018).

\bibitem{magnon_trap} T. Yu, H. C. Wang, M. A. Sentef, H. M. Yu, and G. E. W. Bauer, Phys. Rev. B \textbf{102}, 054429 (2020).
\bibitem{Canming} Y. P. Wang, J. W. Rao, Y. Yang, P. C. Xu, Y. S. Gui, B. M. Yao, J. Q. You, and C.-M. Hu, Phys. Rev. Lett. \textbf{123}, 127202 (2019).


\bibitem{SAW_book2} G. S Kino, {\it Acoustic Waves: Devices, Imaging, And
Analog Signal Processing} (Prentice-Hall, New Jersey,
1987).

 \bibitem {Abrikosov}A. A. Abrikosov, L. P. Gorkov, and I. E. Dzyaloshinski,
\textit{Methods of Quantum Field Theory in Statistical Physics} (Prentice
Hall, Englewood Cliffs, N. J., 1963).

\bibitem {Fetter}A. L. Fetter and J. D. Walecka, \textit{Quantum Theory of
	Many Particle Systems} (McGraw-Hill, New York, 1971).

\bibitem {Mahan}G. D. Mahan, \textit{Many Particle Physics} (Plenum, New York, 1990).

\bibitem {Haug}H. Haug and A. P. Jauho, \textit{Quantum Kinetics in Transport
	and Optics of Semiconductors} (Springer, Berlin, 1996).

\bibitem{Jilei_Ni} J. Chen, C. Liu, T. Liu, Y. Xiao, K. Xia, G. E. W. Bauer, M. Wu, and H. Yu, Phys. Rev. Lett. \textbf{120}, 217202 (2018).

\bibitem{Landau} L. D. Landau and E. M. Lifshitz, \textit{Electrodynamics of Continuous Media}, 2nd ed. (Butterworth-Heinenann,
Oxford, 1984).

\bibitem{Kittel_old} C. Kittel, Phys. Rev. \textbf{110}, 836 (1958).

\bibitem{parameters} Z. Tian, D. Sander, and J. Kirschner, Phys. Rev. B \textbf{79}, 024432 (2009).





\bibitem{Maekawa} S. Maekawa and M. Tachiki,
AIP Conf. Proc. \textbf{29}, 542 (1976).

\bibitem{magnetorotation_1997}D. A. Garanin and E. M. Chudnovsky, Phys. Rev. B \textbf{56},
11102 (1997).

\bibitem{scattering_PRE} Y. Xu, Y. Li, R. K. Lee, and A. Yariv,  Phys. Rev. E {\bf 62}, 7389 (2000).
\bibitem{scattering_PRB} S. Fan, P. R. Villeneuve, and J. D. Jaonnopoulos,  Phys. Rev. B {\bf 59}, 15 882 (1999).



\bibitem {chiral_emitter}C. A. Downing, J. C. L\'opez Carre\~no, F. P. Laussy, E. del Valle, and A. I. Fern\'andez-Dom\'{\i}nguez, Phys. Rev. Lett. {\bf 122}, 057401 (2019).

\bibitem{Gardiner_equivalence} C. W. Gardiner and M. J. Collett, Phys. Rev. A {\bf 31}, 3761 (1985).
\bibitem{Gardiner_chiral} C. W. Gardiner, Phys. Rev. Lett. {\bf 70}, 2269 (1993).

    
\bibitem{gain2} L. Ge, Y. D. Chong, and A. D. Stone, Phys. Rev. A \textbf{85}, 023802 (2012).
\bibitem{gain1}A. Mostafazadeh, J. Phys. A: Math. Theor. \textbf{47}, 505303 (2014).
\bibitem{gain3} A. Galda and V. M. Vinokur, Phys. Rev. B \textbf{94}, 020408(R) (2016).
\bibitem{Peng_gain} Y. S. Cao and P. Yan, arXiv:2006.16510.

\bibitem{PRApplied} A. H.-M\'inguez, F. Maci\'a, J. M. Hern\'andez, J. Herfort, and P. V. Santos, Phys. Rev. Applied \textbf{13}, 044018 (2020).

\bibitem{megahertz1} E. Dieulesaint and D. Royer, \textit{Elastic Waves in Solids
II: Generation, Acousto-Optic Interaction, Applications} (Springer, New York, 2000).
\bibitem{megahertz2} P. Ventura, M. Solal, P. Du lie, J. M. Hode, and F.
Roux, Proceedings of IEEE Ultrasonics Symposium \textbf{1},
1 (1994).

\bibitem{DMI2} R. Verba, I. Lisenkov, I. Krivorotov, V. Tiberkevich, and A. Slavin, Phys. Rev. Appl. \textbf{9}, 064014 (2018).
\bibitem{DMI3} R. Verba, V. Tiberkevich, and A. Slavin, Phys. Rev. Appl. \textbf{12}, 054061 (2019).

\bibitem{referee1}M. Langer, R. A. Gallardo, T. Schneider, S. Stienen, A. Rold\'an-Molina, Y. Yuan, K. Lenz, J. Lindner, P. Landeros, and J. Fassbender, Phys. Rev. B \textbf{99}, 024426 (2019).
\bibitem{referee2}M. Mruczkiewicz, M. Krawczyk, V. K. Sakharov, Yu. V. Khivintsev, Yu. A. Filimonov, and S. A. Nikitov, J. Appl. Phys. \textbf{113}, 093908 (2013).
\bibitem{referee3}S. M. Kukhtaruk, A. W. Rushforth, F. Godejohann, A. V. Scherbakov, and M. Bayer, arXiv:2006.14394.

\end{thebibliography}
\end{document}